\documentclass[12pt]{article}
\pdfoutput=1
\usepackage{pstricks}
\usepackage{color}
\usepackage{amssymb,amsmath,bm,bbm}
\usepackage{epsf}
\usepackage{epsfig}
\usepackage{afterpage}
\usepackage{longtable}
\usepackage{cite}
\usepackage{latexsym,mathrsfs,dsfont}
\usepackage{graphics}
\usepackage{url}
\usepackage{paralist}
\usepackage{bbold}

\newcommand{\e}{\mathrm{e}}

\newcommand{\ord}{\mathcal{O}}

\newcommand{\IM}{\rm{Im}}
\newcommand{\RE}{\rm{Re}}

\newcommand{\tev}{\, {\rm TeV}}
\newcommand{\gev}{\, {\rm GeV}}
\newcommand{\mev}{\, {\rm MeV}}

\newcommand{\vcb}{|V_{cb}|}
\newcommand{\vtd}{|V_{td}|}
\newcommand{\vub}{|V_{ub}|}
\newcommand{\vts}{|V_{ts}|}

\newcommand{\beq}{\begin{equation}}
\newcommand{\eeq}{\end{equation}}
\newcommand{\be}{\begin{equation}}
\newcommand{\ee}{\end{equation}}
\newcommand{\bi}{\begin{itemize}}
\newcommand{\ei}{\end{itemize}}
\newcommand{\ba}{\begin{array}}
\newcommand{\ea}{\end{array}}
\newcommand{\beqa}{\begin{eqnarray}}
\newcommand{\eeqa}{\end{eqnarray}}
\newcommand{\bea}{\begin{eqnarray}}
\newcommand{\eea}{\end{eqnarray}}
\newcommand{\beqn}{\begin{eqnarray}}
\newcommand{\eeqn}{\end{eqnarray}}

\newcommand{\D}{\Delta}

\newcommand{\eps}{\epsilon}

\definecolor{red}{cmyk}{0,1,1,0.4}

\def\kpn{K^+\rightarrow\pi^+\nu\bar\nu}
\def\klpn{K_{L}\rightarrow\pi^0\nu\bar\nu}

\usepackage{fancyhdr}
\advance \headheight by 3.0truept       

\begin{document}

\begin{flushright}
    {FLAVOUR(267104)-ERC-55}\\
    {BARI-TH/13-681}
\end{flushright}

\medskip

\begin{center}
{\LARGE\bf
\boldmath{331 models facing new $b\to s\mu^+\mu^-$ data
}}\\[0.8 cm]
{\bf Andrzej~J.~Buras$^{a,b}$, Fulvia~De~Fazio$^{c}$ and
Jennifer Girrbach$^{a,b}$
 \\[0.5 cm]}
{\small
$^a$TUM Institute for Advanced Study, Lichtenbergstr. 2a, D-85747 Garching, Germany\\
$^b$Physik Department, Technische Universit\"at M\"unchen,
James-Franck-Stra{\ss}e, \\D-85747 Garching, Germany\\
$^c$Istituto Nazionale di Fisica Nucleare, Sezione di Bari, Via Orabona 4,
I-70126 Bari, Italy}
\end{center}

\vskip0.41cm


\abstract{%
\noindent
We investigate how 
the 331 models, based on the gauge group $SU(3)_C\times SU(3)_L\times U(1)_X$
face  new data on $B_{s,d}\to \mu^+\mu^-$ and $B_d\to K^*(K)\mu^+\mu^-$
taking into account present constraints from $\Delta F=2$ observables,
low energy precision measurements, LEP-II and the LHC data.
In these models new sources of flavour and CP violation originate dominantly through  flavour violating interactions of ordinary quarks
and leptons with a new heavy $Z^\prime$ gauge boson. The strength of the relevant couplings is governed by four new parameters  in the
quark
sector and the parameter $\beta$ which in these models determines the charges of new heavy fermions and gauge bosons. We study the
implications of these models  for $\beta=\pm n/\sqrt{3}$ with $n=1,2,3$.  The case $\beta=-\sqrt{3}$
leading to Landau singularities for  $M_{Z^\prime}\approx 4\tev$ can be ruled out when the present constraints on
$Z^\prime$ couplings, in particular from LEP-II, are taken into account.
 For $n=1,2$ interesting results are found for  $M_{Z^\prime}< 4\tev$ with
largest NP effects for $\beta <0$ in   $B_d\to K^*\mu^+\mu^-$ and the ones
in  $B_{s,d}\to\mu^+\mu^-$ for $\beta>0$. As $\RE(C_9^{\rm NP})$  can
reach the values $-0.8$ and  $-0.4$  for $n=2$ and $n=1$, respectively the $B_d\to K^*\mu^+\mu^-$ anomalies can be
softened with
the size depending on
$\Delta M_{s}/(\Delta M_{s})_{\rm SM}$ and the CP-asymmetry $S_{\psi\phi}$.
A correlation between  $\RE(C^{\rm NP}_{9})$ and $\overline{\mathcal{B}}(B_{s}\to\mu^+\mu^-)$, identified for $\beta<0$, implies for {\it 
negative}
$\RE(C^{\rm NP}_{9})$ uniquely suppression of $\overline{\mathcal{B}}(B_{s}\to\mu^+\mu^-)$  relative to its SM value which is favoured by 
the data. In
turn also
$S_{\psi\phi}< S_{\psi\phi}^{\rm SM}$ is favoured with $S_{\psi\phi}$
having dominantly opposite sign to $S_{\psi\phi}^{\rm SM}$
 and closer to its central experimental value. 
 Another triple correlation is the one between $\RE(C^{\rm NP}_9)$, $\overline{\mathcal{B}}(B_{s}\to\mu^+\mu^-)$ and  $\mathcal{B}(B_d\to 
K\mu^+\mu^-)$.
 NP effects in $b\to s\nu\bar\nu$ transitions, $\kpn$
and $\klpn$ turn out  to be small. We
find that the absence of $B_d\to K^*\mu^+\mu^-$ anomalies in the future
data and confirmation of the suppression of $\overline{\mathcal{B}}(B_{s}\to\mu^+\mu^-)$
relative to its SM value would favour $\beta=1/\sqrt{3}$ and $M_{Z^\prime}\approx
3\tev$. Assuming lepton universality,
 we find an upper bound  $|C^{\rm NP}_{9}|\le 1.1 (1.4)$ from
LEP-II data for {\it all}  $Z^\prime$ models with only left-handed flavour
violating couplings to quarks when NP contributions to
$\Delta M_s$ at the level of  $10\%(15\%)$  are allowed.
}

\thispagestyle{empty}
\newpage
\setcounter{page}{1}

\section{Introduction}
The great expectations to find New Physics (NP) at the LHC did not materialize
until now. In particular the order of magnitude enhancements of the branching
ratio for $B_s\to \mu^+\mu^-$ decay over its Standard Model (SM) value, possible in supersymmetric models and models with
tree-level heavy neutral scalar and pseudoscalar exchanges, are presently ruled out.
This is also the case of $\ord(1)$ values of the CP-asymmetry $S_{\psi\phi}$ which could also be accommodated in these models.  A
recent review can be found in  \cite{Buras:2013ooa}.

While for the models in question
these new flavour data are a big disappointment, for other models like the
ones with constrained minimal flavour violation (CMFV), $331$ models \cite{Buras:2012dp} and
 Littlest Higgs Model
with T-parity \cite{Blanke:2009am} they brought a relief as in these
models NP effects were naturally predicted to be small.
On the other hand the most recent data from LHCb and CMS bring new challenges
for the latter models:
\begin{itemize}
\item
The LHCb and CMS collaborations presented new results on $B_{s,d}\to\mu^+\mu^-$
\cite{Aaij:2013aka,Chatrchyan:2013bka,CMS-PAS-BPH-13-007}.  While the
branching ratio for $B_s\to\mu^+\mu^-$ as stated above
turns out to be rather close to the SM prediction, although a bit lower than the latter,
the central value for the one of $B_d\to\mu^+\mu^-$ is by a factor of 3.5 higher  than its SM value.
\item
LHCb collaboration reported new results on angular observables in
$B_d\to K^*\mu^+\mu^-$ that show  departures from SM expectations
\cite{Aaij:2013iag,Aaij:2013qta}. Moreover, new data on the observable $F_L$,
consistent with LHCb value in \cite{Aaij:2013iag} have been presented by
CMS \cite{Chatrchyan:2013cda}.
\end{itemize}

In particular the anomalies in $B_d\to K^*\mu^+\mu^-$ triggered
 two sophisticated analyses \cite{Descotes-Genon:2013wba,Altmannshofer:2013foa}
with the goal to understand
the data and to indicate what type of NP could be responsible
for these departures from the SM. Subsequently several other analyses
of these data have been presented in \cite{Gauld:2013qba,Buras:2013qja,Gauld:2013qja,Beaujean:2013soa,Datta:2013kja,Horgan:2013pva} and very
recently in
 \cite{Descotes-Genon:2013zva}.

The outcome of these efforts can be summarized briefly as follows. There seems
to be a consensus among different groups that NP definitely affects the
Wilson coefficient $C_9$  \cite{Descotes-Genon:2013wba,Altmannshofer:2013foa,Buras:2013qja,Horgan:2013pva,Descotes-Genon:2013zva} with the
value of the shift in $C_9$ depending on the analysis considered:
\be
-1.9\le C^{\rm NP}_{9}\le -0.5.
\ee
There is also a consensus that small negative NP contributions to the
the Wilson coefficient $C_{7\gamma}$ could together with  $C^{\rm NP}_{9}$
provide the explanation of the data \cite{Descotes-Genon:2013wba,Altmannshofer:2013foa}. On the other hand as seen in the analyses in
\cite{Altmannshofer:2013foa,Buras:2013qja,Horgan:2013pva}, a particularly successful scenario is the
one with participation of right-handed currents 
\be
C^{\rm NP}_{9} < 0, \qquad C^{\prime}_{9} >0, \qquad   C^{\prime}_{9}\approx -C^{\rm NP}_{9}.
\ee
However a very recent analysis in \cite{Descotes-Genon:2013zva} challenges
this solution favouring the one with   NP contributions dominantly
represented by $C^{\rm NP}_{9}\approx -1.5$ with much smaller NP contributions
to the remaining Wilson coefficients, in particular $C^{\prime}_{9}$.

For the models  presented here sorting out these differences is  important as
in these models $C^{\prime}_{9}=0$ and as demonstrated in
\cite{Buras:2012dp,Buras:2012jb} NP contributions to  $C_{7\gamma}$ are totally
negligible. Thus $C^{\rm NP}_{9}$ remains the only coefficient which could
help in explaining the $B_d\to K^*\mu^+\mu^-$ anomalies. In our view this
is certainly not excluded
\cite{Descotes-Genon:2013wba,Gauld:2013qba,Buras:2013qja,Gauld:2013qja,Beaujean:2013soa,Descotes-Genon:2013zva}, in particular if these
anomalies would
soften with time.
It should be emphasized at this point that these analyses are subject
to theoretical uncertainties, which have been discussed at length in
\cite{Khodjamirian:2010vf,Beylich:2011aq,Matias:2012qz,Jager:2012uw,Descotes-Genon:2013wba,Hambrock:2013zya} and it remains to be seen
whether the observed anomalies are only
result of statistical fluctuations and/or underestimated error uncertainties.

Assuming that indeed NP is at work here, one of the physical mechanisms behind these deviations that seems to emerge from these studies is
the presence of tree-level $Z^\prime$ exchanges. In \cite{Buras:2012jb} we have presented an anatomy of $Z^\prime$ contributions to flavour
changing neutral current processes (FCNC) identifying various correlations
between various observables characteristic for this NP scenario. Recently
we have analyzed how this scenario faces the new data listed above \cite{Buras:2013qja} including the correlation with the values of
 $C_{B_q}=\Delta M_{q}/(\Delta M_{q})_{\rm SM}$,
$S_{\psi\phi}$ and $S_{\psi K_S}$ which should be precisely determined in  this decade.

The dominant role in \cite{Buras:2013qja} was played by the so-called LHS scenario in which the flavour violating
couplings of $Z^\prime$ to quarks were purely left-handed.
While, in agreement with \cite{Altmannshofer:2013foa} and recently with
\cite{Horgan:2013pva} it has been found that the presence of right-handed couplings leading to a non-vanishing $C_9^\prime$ gives a better
description of the data than the LHS scenario, it is clear that in view of theoretical and experimental
uncertainties the LHS scenario remains as a viable alternative.

The nice virtue of the LHS scenario is that for certain choices
of the $Z^\prime$ couplings the model resembles the structure of CMFV or models with $U(2)^3$ flavour symmetry. Moreover as no new operators
beyond those
present in the SM are present, the non-perturbative uncertainties are the
same as in the SM, still allowing for non-MFV contributions beyond those present in $U(2)^3$ models. In particular the stringent CMFV
relation between
$\Delta M_{s,d}$ and $\mathcal{B}(B_{s,d}\to\mu^+\mu^-)$
\cite{Buras:2003td} valid in the simplest $U(2)^3$ models is violated
in the LHS scenario as analyzed in detail in \cite{Buras:2013qja}.
Another virtue of the LHS scenario is the paucity of its parameters
that enter all flavour observables in a given meson system which should be
contrasted with most NP scenarios outside the MFV framework.
 Indeed, if we concentrate
on $B_s^0-\bar B_s^0$ mixing, $b\to s\mu^+\mu^-$ and $b\to s \nu\bar\nu$
observables, for a given mass $M_{Z^\prime}$ there are only four new parameters to our disposal: the
three couplings (our normalizations of couplings are given
in Section~\ref{sec:2})
\be\label{Step3a}
\Delta_L^{sb}(Z'), \quad \Delta_A^{\mu\bar\mu}(Z'),
\quad \Delta_V^{\mu\bar\mu}(Z'),
\ee
of which the first one is generally complex
and the other two real.
The couplings $\Delta_{A,V}^{\mu\bar\mu}(Z')$ are defined
in (\ref{DeltasVA}) and due to $SU(2)_L$ symmetry implying in LHS
$\Delta_{L}^{\nu\bar\nu}(Z')=\Delta_{L}^{\mu\bar\mu}(Z')$ one also has
\be\label{SU2}
\Delta_{L}^{\nu\bar\nu}(Z')=\frac{\Delta_V^{\mu\bar\mu}(Z')-\Delta_A^{\mu\bar\mu}(Z')}{2}.
\ee
Extending these considerations to $B_d$ and $K$ meson systems brings in
four additional parameters, the complex couplings:
\be
\Delta_L^{db}(Z'), \qquad \Delta_L^{sd}(Z').
\ee

Thus in this general LHS scenario we deal with eight new parameters. Further
reduction of parameters is only possible in a concrete dynamical model.
In this context an interesting class of dynamical models representing LHS scenario are
the 331 models based on the gauge group $SU(3)_C\times SU(3)_L\times U(1)_X$
  \cite{Pisano:1991ee,Frampton:1992wt}. A detailed analysis of FCNC processes
in one of these models has been presented by us in \cite{Buras:2012dp}. Selection of earlier analyses of various aspects of these models
related to our
paper can be found in
\cite{Ng:1992st,Diaz:2003dk,Liu:1993gy,Diaz:2004fs,Liu:1994rx,Rodriguez:2004mw,Promberger:2007py,Agrawal:1995vp,CarcamoHernandez:2005ka,Promberger:2008xg}.

The nice feature of these models is a small number of free parameters which
is lower than present in the general LHS scenario considered
in \cite{Buras:2012jb,Buras:2013qja}. This allows to find certain
correlations between different  meson systems which is not possible in the general case. Indeed
the strength of the relevant $Z^\prime$ couplings  to down-quarks is governed by two mixing parameters, two CP-violating phases and the
parameter $\beta$ which
defines\footnote{ Up to the choice of the representation of the gauge group
according to which the fermions should transform, as will be better
clarified in the next section.} a given 331 model and
determines the charges of new heavy fermions and gauge bosons.
Thus
for a given $M_{Z^\prime}$ and $\beta$ there are only four free parameters to our disposal.
In particular for a given $\beta$, the couplings of $Z^\prime$ to leptons are
fixed. As evident from the general analysis of LHS scenario in
\cite{Buras:2013qja}, knowing the
latter couplings simplifies the analysis significantly, increasing simultaneously the predictive power of the theory.

In \cite{Buras:2012dp} the relevant couplings have been presented for arbitrary
$\beta$ but the detailed FCNC analysis has been only performed for
$\beta=1/\sqrt{3}$. While this model provides interesting results for
$B_{s,d}\to \mu^+\mu^-$, it fails in the case of anomalies in $B_d\to K^*\mu^+\mu^-$ because in this model the coupling
$\Delta_V^{\mu\bar\mu}(Z^\prime)$ and consequently
the Wilson coefficient $C^{\rm NP}_{9}$ turn out to be very small.

It has been pointed out recently in \cite{Gauld:2013qja} that  for $\beta=-\sqrt{3}$  a very different
picture arises. Indeed in this case $\Delta_V^{\mu\bar\mu}(Z^\prime)$
is much larger than for $\beta=1/\sqrt{3}$ so that the $B_d\to K^*\mu^+\mu^-$
anomaly can be in principle successfully addressed. Simultaneously the coupling
 $\Delta_A^{\mu\bar\mu}(Z^\prime)$ turns out to be small so that NP contributions to
$B_s\to\mu^+\mu^-$ are small in agreement with the data.
Moreover aligning the new
mixing matrix $V_L$ with the CKM matrix, the authors end up with a very
simple model in which the only new parameter relevant for their analysis is  $M_{Z^\prime}$
and the negative sign of $C^{\rm NP}_{9}$ required by the $B_d\to K^*\mu^+\mu^-$
anomaly is uniquely predicted.

 Unfortunately this model has several problems, in particular in
the MFV limit considered in
\cite{Gauld:2013qja},
which in our view eliminates it
as a valid description of the present flavour data.
As discussed in Appendix~\ref{beta3} these problems originate in the known 
fact 
that the 331 models with $\beta=\pm\sqrt{3}$
imply a Landau singularity for $\sin^2\theta_W=0.25$ and this value is
reached through the renormalization group evolution of the SM couplings
for $M_{Z^\prime}$ typically around  $4\tev$, scales not much higher than 
the present lower bounds on $M_{Z^\prime}$. 

Yet, the observation of the authors of \cite{Gauld:2013qja} that negative values of $\beta$ could provide solution to $B_d\to
K^*\mu^+\mu^-$ anomalies  motivates us
to generalize our phenomenological analysis of 331 model in  \cite{Buras:2012dp} from $\beta=1/\sqrt{3}$ to arbitrary values of
$\beta$, both positive and negative, for which the Landau singularity in question is avoided up to the very
high scales, even as high as GUTs scales. This generalization is in fact
straight forward as  in  \cite{Buras:2012dp} we have provided formulae for
the $Z^\prime$ couplings to quarks and leptons for arbitrary $\beta$\footnote{ See also \cite{Diaz:2004fs}.} and
the expressions for various flavour observables as functions of $\beta$ can
be directly obtained from the formulae of that paper. In this context
we will concentrate  our analysis on the cases $\beta=\pm n/\sqrt{3}$ with $n=1,2$
choosing $M_{Z^\prime}=3\tev$ in order to satisfy existing bounds from flavour conserving
observables. A simple scaling
law allows  then to obtain predictions for other values of  $M_{Z^\prime}$.

However, in contrast to our numerical analysis in  \cite{Buras:2012dp} which
assumed certain fixed values of $\sqrt{\hat B_{B_s}}F_{B_s}$ and $\sqrt{\hat B_{B_d}}F_{B_d}$ we will investigate in the spirit
of our recent paper
\cite{Buras:2013qja} how our results depend on
\be\label{CBq}
C_{B_q}=\frac{\Delta M_{q}}{(\Delta M_{q})_{\rm SM}}, \qquad
S_{\psi\phi},\qquad  S_{\psi K_S}
\ee
 which should be precisely determined in  this decade.

Our paper is organized as follows. In Section~\ref{sec:2} we review
very briefly the basic aspects of 331 models, recalling their free parameters
and  the general formulae for the couplings of $Z^\prime$ to quarks and leptons
for arbitrary $\beta$. We also present a table with the values of flavour
diagonal couplings of $Z^\prime$ to quarks and leptons for $n=1,2,3$ which
should facilitate other researchers to test precisely these models in
processes not considered by us.
In Section~\ref{sec:3} we
collect  formulae for various Wilson coefficients and one-loop master
functions in terms of the couplings of Section~\ref{sec:2}. This will
allow us to identify certain properties and correlations between various
observables that will be explicitly seen in our numerical analysis.
 As all relevant
formulae for various branching ratios and other observables have been
presented in \cite{Buras:2012dp} we recall in Section~\ref{sec:3a}
only crucial observables and their status in the SM and experiment.
The strategy for our analysis is presented in Section~\ref{sec:4} and
its execution in Section~\ref{sec:5}. In Section~\ref{sec:6} we present
predictions for low energy precision observables which could provide additional tests of the models considered and analyse also
the bounds from LEP-II.  We also comment on the bounds on  $M_{Z^\prime}$ from the LHC. We summarize the main results of our
paper in Section~\ref{sec:7}.  Some useful information can also be found in 
three appendices.

\section{The 331 models and their couplings}\label{sec:2}
\subsection{The 331 models}
The name 331 encompasses a class of models based on the gauge group $SU(3)_C \times SU(3)_L \times U(1)_X$
\cite{Pisano:1991ee,Frampton:1992wt}, that is at first spontaneously broken to the SM   group $SU(3)_C \times SU(2)_L \times
U(1)_Y$ and then  undergoes the spontaneous symmetry breaking to $SU(3)_C \times U(1)_Q$. The extension of the gauge group with
respect to SM leads to interesting consequences.
The first one is that the requirement of anomaly cancellation together with that of asymptotic freedom of QCD implies that the
number of generations must necessarily be equal to the number of colours, hence giving an explanation for the existence of three
generations.
Furthermore, quark generations should transform differently under the action of $SU(3)_L$. In particular, two quark generations
should transform as triplets, one as an antitriplet. Choosing the latter to be the third generation, this different treatment
could be at the origin of the large top quark mass.
This choice imposes that the leptons should transform as antitriplets.
 However, one could choose a different scenario in which the role of
triplets and antitriplets is exchanged, provided that the number of
triplets equals that of antitriplets, in order to fulfil the anomaly
cancellation requirement. Therefore, different versions of the model are obtained according to the way one fixes the fermion 
representations. The fermion
representations for specific 331 models analyzed in our paper are described
in detail in \cite{Buras:2012dp}.

A fundamental relation holds among some of the generators of the group:
\be
Q=T_3+\beta T_8+X,
\ee
where $Q$ indicates the electric charge, $T_3$ and $T_8$ are two of the $SU(3)$ generators and $X$ is the generator of $U(1)_X$.
$\beta$ is a key parameter that defines a specific variant of the model.
The 331 models comprise  several new particles. There are new gauge bosons $Y$ and $V$ and new heavy fermions, all with  electric
charges depending on $\beta$.
Also the Higgs system is extended.

As analyzed in detail in \cite{Diaz:2004fs} and stated in that paper $\beta$ can be arbitrary. Yet due to the fact that in 331
models
\be
M^2_{Z^\prime} = \frac{g^2 u^2 c_W^2}{3[1-(1+\beta^2) s_W^2]} \label{mZprimefin} \,\,
\ee
where $u$ is the vacuum expectation value related to the first symmetry breaking it
is evident that only values of $\beta$ satisfying
\be\label{betacondition}
[1-(1+\beta^2) s_W^2]> 0
\ee
are allowed. With the known value of $s_W^2$ this means that
\be\label{betac1}
|\beta|\le \sqrt{3}
\ee
and in fact the only explicit models analyzed in the literature are the
ones with $\beta=\pm 1/\sqrt{3}$ and $\beta=\pm \sqrt{3}$.
But only for $\beta=\pm 1/\sqrt{3}$ one can avoid the presence of exotic charges both in the fermion and gauge boson sectors. If
one considers only
\be\label{nbeta}
\beta=\pm \frac{n}{\sqrt{3}}, \qquad n=1,2,3
\ee
then for $n=1$  there are
 singly charged $Y^\pm$ bosons and  neutral ones $V^0 (\bar V^0)$, while
for $n=3$ one finds instead two new singly charged bosons $V^\pm$ and two doubly charged ones $Y^{\pm\pm}$. For $n=2$ exotic
charges $\pm 1/2$ and $\pm 3/2$ for
gauge bosons are found. From Table~1 in \cite{Buras:2012dp} we also find
that while for $n=1$ no exotic charges for heavy fermions are present, for $n=2$
heavy quarks carry exotic electric charges $\pm 5/6$ and $\pm 7/6$ while heavy leptons $\pm 1/2$ and $\pm 3/2$. Discovering such
fermions at the LHC would be a spectacular
event. We refer to \cite{Buras:2012dp} for further details.
In principle $\beta$ could be a continuous variable satisfying (\ref{betac1})
but in the present paper we will only consider the cases $n=1,2,3$.

Most importantly for our paper for all $\beta$  a new neutral gauge boson $Z^\prime$ is present. This represents a very appealing
feature, since $Z^\prime$ mediates tree level flavour changing neutral currents (FCNC) in the quark sector and
could be responsible for the recent anomalies as indicated by their recent
extensive analyses.

As in the SM, quark mass eigenstates are defined upon rotation of flavour eigenstates through two unitary matrices $U_L$ (for
up-type quarks) and $V_L$ (for down-type quarks). The relation
\be\label{ULVL}
V_{\rm CKM}=U_L^\dagger V_L
\ee
 holds in analogy with the SM case. However, while in  the SM $V_\text{CKM}$ appears only in charged current interactions and the
two rotation
matrices never appear individually, this is not the case  in this model and both $U_L$ and $V_L$ can
generate tree-level FCNCs mediated by  $Z^\prime$ in the up-quark and
down-quark sector, respectively. But these two matrices have to satisfy the
relation (\ref{ULVL}). A useful parametrization for $V_L$ which we have used
in \cite{Buras:2012dp} is
\begin{equation}
V_L=\left(\begin{array}{ccc}
{\tilde c}_{12}{\tilde c}_{13} & {\tilde s}_{12}{\tilde c}_{23} e^{i \delta_3}-{\tilde c}_{12} {\tilde s}_{13} {\tilde
s}_{23}e^{i(\delta_1
-\delta_2)} & {\tilde c}_{12}{\tilde c}_{23} {\tilde s}_{13} e^{i \delta_1}+ {\tilde s}_{12} {\tilde
s}_{23}e^{i(\delta_2+\delta_3)} \\
-{\tilde c}_{13} {\tilde s}_{12}e^{-i\delta_3} & {\tilde c}_{12}{\tilde c}_{23} + {\tilde s}_{12}
 {\tilde s}_{13} {\tilde s}_{23}e^{i(\delta_1-\delta_2-\delta_3)} & - {\tilde s}_{12} {\tilde s}_{13}{\tilde c}_{23}e^{i(\delta_1
-\delta_3)}
-{\tilde c}_{12} {\tilde s}_{23} e^{i \delta_2} \\
- {\tilde s}_{13}e^{-i\delta_1} & -{\tilde c}_{13} {\tilde s}_{23}e^{-i\delta_2} & {\tilde c}_{13}{\tilde c}_{23}
\end{array}\right) \,\,.\label{VL-param}
\end{equation}

This matrix implies through (\ref{ULVL}) new sources of flavour violation
in the up-sector. However, when $U_L=\mathds{1}$ as used in \cite{Gauld:2013qja} $V_L=V_{\rm CKM}$ and we deal with a particular simple
CMFV model.

With this parametrization, the $Z^\prime$ couplings to quarks, for the three meson systems,  $K$, $B_d$ and
$B_s$
\be\label{Deltas}
\Delta^{sd}_L(Z'), \qquad \Delta^{bd}_L(Z') \qquad \Delta^{bs}_L(Z')
\ee
depend only on four new parameters (explicit formulae are given in \cite{Buras:2012dp}):
\be\label{newpar}
\tilde s_{13}, \quad \tilde s_{23}, \quad \delta_1, \quad \delta_2
\ee
with $\tilde s_{13}$ and $\tilde s_{23}$ being positive definite and $\delta_i$ in
the range $[0,2\pi]$.
Therefore for fixed  $M_{Z'}$ and $\beta$, the $Z'$ contributions to all processes
analyzed by us depend only on these parameters implying very strong
correlations between NP effects to various observables.
Indeed, as seen in (\ref{VL-param})
the $B_d$ system involves only the parameters ${\tilde s}_{13}$ and $\delta_1$ while the $B_s$ system depends  on
${\tilde s}_{23}$ and $\delta_2$. Moreover, stringent correlations between observables in $B_{d,s}$ sectors and in the kaon
sector are found since kaon physics depends on ${\tilde s}_{13}$, ${\tilde s}_{23}$ and $\delta_2 - \delta_1$.
A very constraining feature of this models is that the diagonal couplings
of $Z^\prime$ to quarks and leptons are fixed for a given $\beta$, except
for a weak dependence on $M_{Z^\prime}$ due to running of $\sin^2\theta_W$ 
{ provided $\beta$ differs significantly from $\pm\sqrt{3}$.}

\subsection{The couplings}
We will now recall those couplings for arbitrary $\beta$ that are relevant
for our paper. The expressions for other couplings and masses of new gauge bosons and fermions as well as expressions for their
electric charges that depend on
$\beta$ can be found in \cite{Buras:2012dp}.

Central for our analysis is the function
\be\label{central}
f(\beta)=\frac{1}{1-(1+\beta^2)s_W^2} > 0
\ee
where the positivity of this function results from the reality of
$M_{Z^\prime}$ as stressed above.

The following properties
should be noted:
\begin{itemize}
\item
For $\beta\approx \sqrt{3}$ there is a Landau singularity for $s_W^2=0.25$. As
at $M_W$ one has $s_W^2\approx 0.23$ (with exact number depending on its definition considered)
and renormalization group evolution of weak couplings increases  $s_W^2$
with increasing scale,  $s_W^2(M_{Z^\prime})$ reaches 0.25 and the singularity
in question for $M_{Z^\prime}\approx 4\tev$.
\item
For $|\beta|\le \sqrt{3}-0.20$ this problem does not arise even up to the GUTs
scales.
\end{itemize}

While we will specifically consider only the cases
 $\beta=\pm n/\sqrt{3}$ with $n=1,2,3$ we list here the formulae
for the relevant couplings for arbitrary real $\beta\not=\sqrt{3}$ satisfying (\ref{betac1}). The case $\beta =\sqrt{3}$ is
considered separately  in
Appendix~\ref{examples}.

 The important point which we would like to make here is that the
     couplings of $Z^\prime$ to quarks and leptons have to be evaluated at the
scale $\mu$ at which $Z^\prime$ is integrated out, that is at
$\mu=\ord(M_{Z^\prime})$ and not at $M_W$. For $n=1$ this difference is irrelevant. For
$n=2$ it plays a role if acceptable precision is required and it is crucial
for $n=3$. The values of couplings listed by us and in Appendix~\ref{examples}
correspond to $\mu=M_{Z^\prime}$ with the latter specified below.

Denoting the elements of the matrix $V_L$ in (\ref{VL-param}) by $v_{ij}$,
the relevant couplings for quarks are then given as follows:
{\allowdisplaybreaks
\begin{subequations}
\begin{align}
 \Delta_L^{ij}(Z') &=\frac{g}{\sqrt{3}}c_W \sqrt{f(\beta)} v_{3i}^*v_{3j}\,,\\
 \Delta_L^{ji}(Z') &=\left[ \Delta_L^{ij}(Z')\right]^\star\,,\qquad \Delta_L^{u\bar u}(Z') =\Delta_L^{d\bar d}(Z') \\
\Delta_L^{d\bar d}(Z') & = \frac{g}{2\sqrt{3}c_W}\sqrt{f(\beta)} \left[-1+(1+\frac{\beta}{\sqrt{3}})s_W^2\right]
\,,\\
\Delta_R^{u\bar u}(Z')& =  \frac{g}{2\sqrt{3}c_W}\sqrt{f(\beta)}\frac{4}{\sqrt{3}}\beta s_W^2=-2 \Delta_R^{d\bar d}(Z')\,,\\
\Delta_V^{d\bar d}(Z') & = \frac{g}{2\sqrt{3}c_W}\sqrt{f(\beta)} \left[-1+(1-\frac{\beta}{\sqrt{3}})s_W^2\right]\,,\\
\Delta_A^{d\bar d}(Z') & = \frac{g}{2\sqrt{3}c_W}\sqrt{f(\beta)} \left[1-(1+\sqrt{3}\beta)s_W^2\right]\,,\\
\Delta_V^{u\bar u}(Z') & = \frac{g}{2\sqrt{3}c_W}\sqrt{f(\beta)} \left[-1+(1+\frac{5}{\sqrt{3}}\beta)s_W^2\right]\,,\\
\Delta_A^{u\bar u}(Z') & = \frac{g}{2\sqrt{3}c_W}\sqrt{f(\beta)} \left[1-(1-\sqrt{3}\beta)s_W^2\right].
\end{align}
\end{subequations}}%
The diagonal couplings are valid for the first two generations of quarks
neglecting the very small non-diagonal contributions in the matrices $V_L$ and
$U_L$. For the third generation there is an additional term which can be found
in \cite{Buras:2012dp}.

For leptons we have
{\allowdisplaybreaks
\begin{subequations}
\begin{align}
\Delta_L^{\nu\bar\nu}(Z') & = \frac{g}{2\sqrt{3}c_W}\sqrt{f(\beta)} \left[1-(1+\sqrt{3}\beta)s_W^2\right]
\,,\\
\Delta_L^{\mu\bar\mu}(Z')&=\Delta_L^{\nu\bar\nu}(Z')\,,\\
\Delta_R^{\mu\bar\mu}(Z')&= \frac{-g~\beta~ s_W^2}{c_W} \sqrt{f(\beta)}\\
 \Delta_V^{\mu\bar\mu}(Z') &=\frac{g}{2\sqrt{3}c_W} \sqrt{f(\beta)}
~\left[1-(1+3\sqrt{3}\beta)s_W^2\right]
 \,,\\
 \Delta_A^{\mu\bar\mu}(Z') &=\frac{g}{2\sqrt{3}c_W} \sqrt{f(\beta)}
~\left[-1+(1-\sqrt{3}\beta)s_W^2\right]
\end{align}
\end{subequations}}%
where we have defined
\begin{align}\label{DeltasVA}
\begin{split}
 &\Delta_V^{\mu\bar\mu}(Z')= \Delta_R^{\mu\bar\mu}(Z')+\Delta_L^{\mu\bar\mu}(Z'),\\
&\Delta_A^{\mu\bar\mu}(Z')= \Delta_R^{\mu\bar\mu}(Z')-\Delta_L^{\mu\bar\mu}(Z').
\end{split}
\end{align}

\begin{figure}[!tb]
 \centering
\includegraphics[width = 0.45\textwidth]{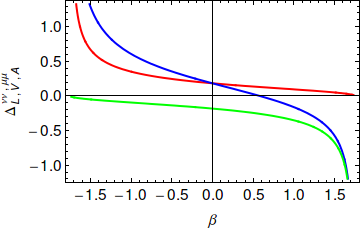}
\caption{ $\Delta_L^{\nu\bar\nu}(Z')$ (red), $\Delta_V^{\mu\bar\mu}(Z')$ (blue)
and $\Delta_A^{\mu\bar\mu}(Z')$ (green) as functions of $\beta$ for $s_W^2=0.249$ and $g=0.633$. This plot applies to $\beta\not=\sqrt{3}$.
}\label{fig:F7}~\\[-2mm]\hrule
\end{figure}

In Fig.~\ref{fig:F7} we show  $\Delta_L^{\nu\bar\nu}(Z')$, $\Delta_V^{\mu\bar\mu}(Z')$
and $\Delta_A^{\mu\bar\mu}(Z')$ as functions of $\beta$ for $s_W^2=0.249$ and
$g=0.633$
corresponding to $M_{Z^\prime}=3\tev$. We observe the
following features:
\begin{itemize}
\item
$\Delta_L^{\nu\bar\nu}(Z')$ and  $\Delta_A^{\mu\bar\mu}(Z')$  have definitive
signs in the full range of $\beta$: {\it positive} and {\it negative}, respectively.
\item
$\Delta_V^{\mu\bar\mu}(Z')$ can have both signs and for fixed $|\beta|$ its magnitude is larger for $\beta < 0$. In fact the models with
negative $\beta$ are then favoured
by the $B_d\to K^*\mu^+\mu^-$ anomalies. As noticed in  \cite{Gauld:2013qja} in this case in the limit of CMFV one automatically
obtains $C^{\rm NP}_9<0$
as required by experiment. If there are new sources of flavour violation
represented by the matrix $V_L$ then the oasis in the space of new parameters
has to  be chosen for which in the case of negative $\beta$ one still has
 $\RE(C^{\rm NP}_9)<0$. This will in turn have consequences for other processes as
we will see below.
\end{itemize}
\subsection{Various 331 models}
It is instructive to list the values of the resulting couplings for the  models with $n=1,2,3$ in (\ref{nbeta}). We do
this for flavour
diagonal couplings in  Table~\ref{tab:diagonal}
for $s_W^2=0.249$ and $g=0.633$ valid at $M_{Z^\prime}=3\tev$ except
for $n=3$, where we use $M_{Z^\prime}=2\tev$ in order not to be too close to
the Landau singularity. In Appendix~\ref{examples}
we give  explicit formulae for these couplings in terms of $\sin^2\theta_W$
as well as expressions for flavour violating couplings. Here and in Appendix~\ref{examples} we
give also $Z$-couplings.

\begin{table*}[!tb]
\centering
\begin{tabular}{|c || c |c |c| c| c|c||c|}\hline
 & \multicolumn{6}{c||} {$\beta$ in $Z^\prime$ couplings  } & $Z$ couplings \\
\hline
 & $1/\sqrt{3}$ & $-1/\sqrt{3}$ & $2/\sqrt{3}$ &  $-2/\sqrt{3}$ &
$\sqrt{3}$  &$-\sqrt{3}$&\\
\hline
\hline
$\Delta_L^{u{\bar u}}$ & -0.172 & -0.215  & -0.191 &  -0.299 & -0.849 &  -1.672 & 0.257
\\
$\Delta_R^{u{\bar u}}$ & 0.086  & -0.086 & 0.216 &  -0.216 &1.645 &  -1.645 & -0.115
\\
$\Delta_V^{u{\bar u}}$ & -0.087 &  -0.301 & 0.026  & -0.515 & 0.796    & -3.316  & 0.143
\\
$\Delta_A^{u{\bar u}}$ & 0.258 & 0.130 & 0.407  &  0.082 &  2.494 & 0.027 & -0.372
\\
\hline \hline
$\Delta_L^{d{\bar d}}$ & -0.172 & -0.215 & -0.191 & -0.299 &  -0.849 &  -1.672 & -0.315
\\
$\Delta_R^{d{\bar d}}$ & -0.043 & 0.043 & -0.108  & 0.108 &  -0.822 & 0.822 & 0.057
\\
$\Delta_V^{d{\bar d}}$ & -0.215  & -0.172 & -0.299  &  -0.191& -1.672 & -0.849 & -0.257
\\
$\Delta_A^{d{\bar d}}$ & 0.130  &0.258 & 0.082  & 0.407 &  0.027& 2.494 & 0.372
\\
\hline \hline
$\Delta_L^{\mu{\bar \mu}}$ & 0.130 & 0.258 & 0.082 & 0.407 & 0.027 & 2.494 & -0.199
\\
$\Delta_R^{\mu{\bar \mu}}$ & -0.128 & 0.128 & -0.324  & 0.324 & 0.054 &2.467 & 0.172
\\
$\Delta_V^{\mu{\bar \mu}}$ & 0.001 & 0.386 & -0.242 & 0.731 & 0.080 & 4.961& -0.028
\\
$\Delta_A^{\mu{\bar \mu}}$ & -0.258 & -0.130 & -0.407 &  -0.082 & 0.027 & -0.027& 0.372
\\
\hline \hline
$\Delta_L^{\nu{\bar \nu}}$ & 0.130 &  0.258 & 0.082 &  0.407 & 0.027 & 2.494 & 0.372
\\
\hline
\end{tabular}
\caption{Diagonal $Z^\prime$ couplings to fermions for different $\beta$ and SM $Z$ couplings to fermions (last column).
 We have used $\sin^2\theta_W=0.249$ for
$\beta=\pm 1/\sqrt{3}$ and $\beta=\pm 2/\sqrt{3}$, $\sin^2\theta_W=0.246$ for  $\beta=\pm\sqrt{3}$ and $\sin^2\theta_W=0.231$ for
$Z$-couplings.
}\label{tab:diagonal}~\\[-2mm]\hrule
\end{table*}

\section{Master formulae for one-loop functions and Wilson coefficients}\label{sec:3}
\subsection{New physics contributions}
We collect here for completeness the corrections to SM one-loop functions and
relevant Wilson coefficients as functions of the couplings listed in the
previous section.

In the case of $\Delta F=2$ transitions  governed by the function $S$ we have
\be\label{Zprime1}
\Delta S(B_q)=
\left[\frac{\Delta_L^{bq}(Z^\prime)}{\lambda_t^{(q)}}\right]^2
\frac{4\tilde r}{M^2_{Z^\prime}g_{\text{SM}}^2}, \qquad
\Delta S(K)=
\left[\frac{\Delta_L^{sd}(Z^\prime)}{\lambda_t^{(K)}}\right]^2
\frac{4\tilde r}{M^2_{Z^\prime}g_{\text{SM}}^2}
\ee
where
\be\label{gsm}
g_{\text{SM}}^2=4\frac{G_F}{\sqrt 2}\frac{\alpha}{2\pi\sin^2\theta_W}\,, \qquad
\lambda^{(K)}_i=V_{is}^*V_{id},\qquad \lambda_t^{(q)}=V_{tb}^*V_{tq}
\ee
and $\tilde r$ is a QCD factor calculated in \cite{Buras:2012dp}.
 One finds $\tilde r=0.965$, $\tilde r=0.953$ and $\tilde r = 0.925$ for $M_{Z^\prime} =2,~3, ~10\tev$, respectively.
It should be remarked that $g_{\text{SM}}^2$
and $\sin^2\theta_W$ appearing outside the $Z^\prime$ couplings, like in~(\ref{C9}) and (\ref{C10}) below, should be evaluated at $M_Z$ with 
input
values given in Table~\ref{tab:input} as they are just related to the
overall normalization of Wilson coefficients and  rescale relative to SM
contributions.

For decays $B_q\to\mu^+\mu^-$ with $q=d,s$  governed by the function $Y$ one has
\be
\Delta Y(B_q)=
\left[\frac{\Delta_{A}^{\mu\bar\mu}(Z')}{M_{Z'}^2g^2_{\rm SM}}\right]
\frac{\Delta_{L}^{qb}(Z')}{ V_{tq}^\ast V_{tb}}
\ee
and for $K_L\to\mu^+\mu^-$
\be
\Delta Y(K)=
\left[\frac{\Delta_{A}^{\mu\bar\mu}(Z')}{M_{Z'}^2g^2_{\rm SM}}\right]
\frac{\Delta_{L}^{sd}(Z')}{ V_{ts}^\ast V_{td}}.
\ee

Similarly for $b\to q\nu\bar\nu$ transitions  governed by the function $X$ one finds
\be\label{DXB}
\Delta X(B_q)=
\left[\frac{\Delta_{L}^{\nu\nu}(Z')}{g^2_{\rm SM}M_{Z'}^2}\right]
\frac{\Delta_{L}^{qb}(Z')}{ V_{tq}^\ast V_{tb}}
\ee
and for $\kpn$ and $\klpn$
\be\label{XLK}
\Delta X(K)=\left[\frac{\Delta_L^{\nu\bar\nu}(Z')}{g^2_{\rm SM}M_{Z'}^2}\right]
                                       \frac{\Delta_L^{sd}(Z')}{V_{ts}^* V_{td}}.
\ee

The corrections from NP to the Wilson coefficients $C_9$ and $C_{10}$ relevant
for $b\to s\mu^+\mu^-$ transitions and used in the recent literature
are given as follows\footnote{The coefficients $C_V$ and $C_A$ in \cite{Buras:2012dp} are obtained by multiplying $C_9$ and
$C_{10}$ by $\sin\theta_W^2$,
respectively.}
\begin{align}
 \sin^2\theta_W C^{\rm NP}_9 &=-\frac{1}{g_{\text{SM}}^2M_{Z^\prime}^2}
\frac{\Delta_L^{sb}(Z')\Delta_V^{\mu\bar\mu}(Z')} {V_{ts}^* V_{tb}} ,\label{C9}\\
   \sin^2\theta_W C^{\rm NP}_{10} &= -\frac{1}{g_{\text{SM}}^2M_{Z^\prime}^2}
\frac{\Delta_L^{sb}(Z')\Delta_A^{\mu\bar\mu}(Z')}{V_{ts}^* V_{tb}}\label{C10}.
 \end{align}

\subsection{Correlations}
These formulae imply certain relations that are useful for the subsequent
sections. First of all we have the ratio
\be\label{AV}
R_1=\frac{ C^{\rm NP}_{9}}{C^{\rm NP}_{10}}=\frac{ \RE(C^{\rm NP}_{9})}{\RE(C^{\rm NP}_{10})}=\frac{ \IM(C^{\rm NP}_{9})}{\IM(C^{\rm NP}_{10})}=\frac{\Delta_V^{\mu\bar\mu}(Z')}{\Delta_A^{\mu\bar\mu}(Z')},
\ee
which involves only leptonic couplings and  depends only on $\beta$.
This ratio is given in Table~\ref{tab:Ri} for different values of $\beta$ and
 $s_W^2=0.249$ except for $\beta=\pm\sqrt{3}$ where we use $s_W^2=0.246$.
We observe that for  $\beta < 0$
these two coefficients are predicted to have opposite signs
independently
of the $Z^\prime$ couplings to quarks and as  $C^{\rm SM}_{10}$ and
$C^{\rm SM}_{9}$ have also opposite signs  (see (\ref{CSM910})).
  $\RE(C^{\rm NP}_{9})$ and NP contributions to $\overline{\mathcal{B}}(B_{s}\to\mu^+\mu^-)$ are correlated with each
other. This means that $\RE(C^{\rm NP}_{9})<0$ required by $B_d\to K^*\mu^+\mu^-$
data implies for $\beta<0$ uniquely suppression of $\overline{\mathcal{B}}(B_{s}\to\mu^+\mu^-)$ relative to its SM value which is favoured 
by
the data.  On the other hand for  $\beta=1/\sqrt{3}$ the ratio  in (\ref{AV}) is tiny and  for $\beta>  1/\sqrt{3}$ it is positive implying 
that
 NP contributions  to $\overline{\mathcal{B}}(B_{s}\to\mu^+\mu^-)$ and
$\RE(C^{\rm NP}_{9})$ are anti-correlated with each other. Consequently in this case
$\RE(C^{\rm NP}_{9})<0$ required by $B_d\to K^*\mu^+\mu^-$ anomaly  implies
the enhancement of $\overline{\mathcal{B}}(B_{s}\to\mu^+\mu^-)$ which is
presently not supported by the data but this could change in the future.
We will see all this explicitly in Section~\ref{sec:5}.

A complementary relation valid in any LHS model that this time does not 
depend on the lepton couplings is  \cite{Buras:2013qja}
\be\label{IMRE}
\frac{\IM(C_9^{\rm NP})}{\RE(C_9^{\rm NP})}=\frac{\IM(C_{10}^{\rm NP})}{\RE(C_{10}^{\rm NP})}=\tan(\delta_{2}-\beta_s).
\ee
Two important points should be noticed here. These two ratios have to be equal to each other.  Moreover they are the same in the two oases resulting from $\Delta F=2$ constraint.

Next the ratios
\be\label{master3}
R_2=
\frac{\Delta Y(B_q)}{\Delta X(B_q)}=\frac{\Delta Y(K)}{\Delta X(K)}=
\frac{\Delta_A^{\mu\bar\mu}(Z^\prime)}{\Delta_L^{\nu\bar\nu}(Z^\prime)}
\ee
express the relative importance of $Z^\prime$ contributions within a given
meson system to decays with muons and neutrinos in the final state. While
investigating the numbers for $R_2$ in Table~\ref{tab:Ri} we should recall that in
the SM $Y\approx 1.0$ while $X\approx 1.5$ which means that it is easier
to make an impact on decays to muons. While from this ratio we cannot conclude
whether a given branching ratio is enhanced or suppressed as the quark couplings cancel in this ratio, the message is clear:
\begin{itemize}
\item
For $\beta>0$ NP effects in decays to muons governed by axial-vector couplings
 are much larger than in decays to neutrinos,
whereas the opposite is true for $\beta<0$. Therefore in the latter case which
is chosen by the $B_d\to K^*\mu^+\mu^-$ anomaly we can expect also measurable
effects in decays to neutrinos.
\item
Very importantly in all models, except $\beta=\sqrt{3}$, considered NP effects in $B_s\to\mu^+\mu^-$
are anti-correlated with the ones in $b\to s\nu\bar\nu$ transitions.
\end{itemize}

\begin{table}[!tb]
\centering
\begin{tabular}{|c||c|c|c|c|c|c|}
\hline
 $\beta$  & $1/\sqrt{3}$ & $-1/\sqrt{3}$ & $2/\sqrt{3}$ &  $-2/\sqrt{3}$ &
$\sqrt{3}$  &$-\sqrt{3}$\\
\hline
\hline
  \parbox[0pt][1.6em][c]{0cm}{} $R_1$ & $-0.004$ & $-2.98$  & $0.59$ & $-8.87$ & $3.0$ &$-185.5$\\
 \parbox[0pt][1.6em][c]{0cm}{} $R_2$ &  $-1.99$ & $-0.50$ & $-4.94$ & $-0.20$ & $1.0$ & $-0.01$\\
 \parbox[0pt][1.6em][c]{0cm}{} $R_3$ & $0.61$  & $1.22$ & $0.39$ & $1.93$ &$0.13$ & $11.8$\\
 \parbox[0pt][1.6em][c]{0cm}{} $R_4$   & $0.67$  & $0.67$ & $0.42$ & $0.42$ &$0.016$ & $0.016$\\
\parbox[0pt][1.6em][c]{0cm}{} $R_5$   & $0.50$  & $1.00$ & $0.25$ & $1.25$ &$0.016$ & $1.49$\\
\parbox[0pt][1.6em][c]{0cm}{} $R_6$   & $-1.00$  & $-0.50$ & $-1.25$ & $-0.25$ &$0.016$ & $-0.016$\\
 \hline
\end{tabular}
\caption{Values of the ratios $R_i$  for different $\beta$
setting $\sin^2\theta_W=0.249$ except for $\beta=\pm\sqrt{3}$ where we use $s_W^2=0.246$.
}\label{tab:Ri}~\\[-2mm]\hrule
\end{table}

Important are also the relations between the $Z'$ contributions to $\Delta F=1$
($X$ and $Y$ functions) and $\Delta F=2$ ($S$ functions) observables. We have
\be\label{master2}
\frac{\Delta X(B_q)}{\sqrt{\Delta S(B_q)^*}}=a_q\frac{\Delta_L^{\nu\bar\nu}(Z')}{2\sqrt{\tilde r}g_{\rm
SM}M_{Z'}}=-a_q\frac{0.085}{\sqrt{\tilde r}}\left(\frac{3\tev}{M_{Z'}}\right) R_3
\ee
where
\be
R_3=
\frac{\left[1-(1+\sqrt{3}\beta)s_W^2\right]}{\sqrt{1-(1+\beta^2)s_W^2}},
\ee
$a_d=1$ and $a_s=-1$.

We also have
\be\label{master4}
\frac{\Delta X(K)}{\sqrt{\Delta S(K)}}=\frac{\Delta X(B_s)}{\sqrt{\Delta S(B_s)^*}}.
\ee

As presently the constraints on 331 models are dominated by $\Delta F=2$ transitions we observe that for a given allowed size of
$\Delta S(B_q)$, NP effects
in the one-loop functions in question are proportional to $1/M_{Z^\prime}$
and this dependence is transfered to branching ratios
in view of the fact that the dominant NP contributions are present there
as interference between SM and NP contributions. That these effects
are only suppressed like  $1/M_{Z^\prime}$  and not like  $1/M^2_{Z^\prime}$ is
the consequence of the increase with $M_{Z^\prime}$ of the allowed values of
the couplings $\Delta_L^{ij}(Z^\prime)$ extracted from $\Delta F=2$
observables, the point already stressed in \cite{Buras:2012dp}. In
summary, denoting by $\Delta\mathcal{O}^{\rm NP}(M_{Z^\prime}^{(i)})$ NP contributions to a given $\Delta F=1$ observable in
$B_s$ and $B_d$ decays
at two ($(i=1,2)$ different values $M_{Z^\prime}^{(i)}$ we have a {\it scaling law}
\be\label{scaling}
\Delta\mathcal{O}^{\rm NP}(M_{Z^\prime}^{(1)})=\frac{M_{Z^\prime}^{(2)}}{M_{Z^\prime}^{(1)}}
\Delta\mathcal{O}^{\rm NP}(M_{Z^\prime}^{(2)}).
\ee
independently of $\beta$ and $C_{B_q}$. However the size of NP effect will
depend on these two parameters as seen already in the case of $\beta$
 in Table~\ref{tab:Ri} and we will see this more explicitly in Section~\ref{sec:5}.

 While this scaling law would apply in the case of the absence of correlations between $B_q$ and $K$ systems also to $K$ decays,
in the 331 models the
situation is different as we will now demonstrate.
Indeed in these models there is a correlation
between the $Z'$ effects in $\Delta F=2$
master functions in different meson systems
\be\label{master1}
\frac{\Delta S(K)}{\Delta S(B_d)\Delta S(B_s)^*}=\frac{M^2_{Z'}g^2_{\rm SM}}{4 \tilde
r}\left[\frac{\Delta_L^{sd}(Z')}{\Delta_L^{bd}(Z')\Delta_L^{bs*}(Z')}\right]^2=
\frac{3.68}{\tilde r}\left(\frac{M_{Z'}}{3\tev}\right)^2 R_4
\ee
where
\be
R_4=1-(1+\beta^2)s_W^2.
\ee
Here and in following equations we set $|V_{tb}|=1$ and
$\tilde c_{13}=\tilde c_{23}=1$ if necessary. As the present data and
lattice results imply $|\Delta S(B_q)|<0.25$  and $R_4<0.7$ in all models,
 NP effects in $\varepsilon_K$ are typically below $10\%$, which is welcome
as with input parameters in Table~\ref{tab:input}  $\varepsilon_K$ within
the SM is in good agreement with the data.

Combining then relations (\ref{master3}), (\ref{master2}), (\ref{master4}) and  (\ref{master1})
we find
\be\label{master6}
\Delta X(K)=\frac{0.16}{\tilde r} R_5\sqrt{\Delta S(B_d)\Delta S(B_s)^*},\qquad R_5=R_3\sqrt{R_4}
\ee
\be\label{master6a}
\Delta Y(K)=\frac{0.16}{\tilde r} R_6\sqrt{\Delta S(B_d)\Delta S(B_s)^*},\qquad R_6=R_2 R_3\sqrt{R_4}
\ee
with the values of $R_5$ and $R_6$ given in Table~\ref{tab:Ri}. We observe that
$\Delta X(K)$ and $\Delta Y(K)$  do not depend on $M_{Z^\prime}$ when the parameters in $V_L$
are constrained through $B^0_{s,d}-\bar B^0_{s,d}$ mixings. This fact has
already been noticed in \cite{Buras:2012dp} but these explicit relations are new.

For the models considered in detail by us $R_5<1.3$ and as $|\Delta S(B_q)|<0.25$ we find that $|\Delta X(K)|\le 0.05$ which
implies a correction to
$\kpn$ and $\klpn$ of at most  $10\%$ at the level of the branching ratios.

As far as
the Wilson coefficients $C_9^{\rm NP}$ and
$C_{10}^{\rm NP}$ are concerned we have two important relations
\be\label{SC9}
(\Delta S(B_q))^*=4\tilde rg^2_{\rm SM} M_{Z^\prime}^2\sin^4\theta_W \left[\frac{C_9^{\rm
NP}}{\Delta_V^{\mu\bar\mu}(Z')}\right]^2=0.327 \left[\frac{C_9^{\rm
NP}}{\Delta_V^{\mu\bar\mu}(Z')}\right]^2 \left[\frac{M_{Z^\prime}}{3\tev}\right]^2,
\ee
\be\label{SC10}
(\Delta S(B_q))^*=4\tilde rg^2_{\rm SM}M_{Z^\prime}^2\sin^4\theta_W \left[\frac{C_{10}^{\rm
NP}}{\Delta_A^{\mu\bar\mu}(Z')}\right]^2=0.327
\left[\frac{C_{10}^{\rm NP}}{\Delta_A^{\mu\bar\mu}(Z')}\right]^2 \left[\frac{M_{Z^\prime}}{3\tev}\right]^2,
\ee
which we have written in a form suitable for the analysis in Section~\ref{sec:5}. We recall that $S_{\rm SM}=S_0(x_t)=2.31$. Both
relations are valid for
$B_s$ and $B_d$ systems as indicated on the l.h.s of these equations
and the Wilson coefficients on the r.h.s should be appropriately adjusted
to the case considered.

The virtue of these relation is their independence of the new parameters in
(\ref{newpar}) so that for a given $\beta$ the  size of
$C_{9}^{\rm NP}$ and $C_{10}^{\rm NP}$ allowed by the $\Delta F=2$ constraints
can be found. In particular in the case of a real $C_{9}^{\rm NP}$ and $C_{10}^{\rm NP}$, corresponding to $V_L=V_{\rm CKM}$,
 $\Delta S(B_q)$ and $\Delta M_q$ will be enhanced which is only allowed
if the SM values of  $\Delta M_q$ will turn out to be below the data.
If this will not be the case the only solution is to misalign $V_L$ and $V_{\rm CKM}$ which results in complex $C_{9}^{\rm NP}$
and $C_{10}^{\rm NP}$ and  consequently novel CP-violating effects.

\section{Crucial observables}\label{sec:3a}
\boldmath
\subsection{$\Delta F=2$ observables}
\unboldmath
The $B_s^0-\bar B_s^0$ observables are fully described in 331 models  by the function
\begin{equation}\label{Seff}
S(B_s)=S_0(x_t)+\Delta S(B_s)\equiv|S(B_s)|e^{-i2\varphi_{B_s}},
\end{equation}
where $S_0(x_t)$ is the real one-loop SM box function and the additional generally complex term has been given in (\ref{Zprime1}).

The two observables of interest, $\Delta M_s$ and $S_{\psi\phi}$ are then
given by
\be\label{DMs}
\Delta M_s =\frac{G_F^2}{6 \pi^2}M_W^2 m_{B_s}|V^*_{tb}V_{ts}|^2   F_{B_s}^2\hat B_{B_s} \eta_B |S(B_s)|\,
\ee
and
\begin{equation}
S_{\psi\phi} =  \sin(2|\beta_s|-2\varphi_{B_s})\,, \qquad  V_{ts}=-\vts e^{-i\beta_s}.
\label{eq:3.44}
\end{equation}
with $\beta_s\simeq -1^\circ\,$.

In the case of $B_d^0$ system the corresponding formulae are obtained from
(\ref{Seff}) and (\ref{DMs}) by replacing $s$ by $d$. Moreover (\ref{eq:3.44})
is replaced by
\begin{equation}
S_{\psi K_S} =  \sin(2\beta-2\varphi_{B_d})\,, \qquad  V_{td}=\vtd e^{-i\beta}.
\label{eq:3.45}
\end{equation}
With the input for $\vub$ and $\vcb$ in Table~\ref{tab:input} and $\gamma=68^\circ$ there is
a good agreement of the SM with data on $S_{\psi K_S}$ and $\varepsilon_K$.

In the SM one has\footnote{The central value of $\vts$ corresponds roughly to the central $\vcb=0.0424$ obtained from tree-level
inclusive
decays \cite{Gambino:2013rza}.}
\begin{equation}\label{DMS}
(\Delta M_{s})_{\rm SM}=
18.8/{\rm ps}\cdot\left[
\frac{\sqrt{\hat B_{B_s}}F_{B_s}}{266\mev}\right]^2
\left[\frac{S_0(x_t)}{2.31}\right]
\left[\frac{\vts}{0.0416} \right]^2
\left[\frac{\eta_B}{0.55}\right] \,,
\end{equation}

\begin{equation}\label{DMD}
(\Delta M_d)_{\rm SM}=
0.54/{\rm ps}\cdot\left[
\frac{\sqrt{\hat B_{B_d}}F_{B_d}}{218\mev}\right]^2
\left[\frac{S_0(x_t)}{2.31}\right]
\left[\frac{\vtd}{8.8\cdot10^{-3}} \right]^2
\left[\frac{\eta_B}{0.55}\right]\,.
\end{equation}
For the central values of the parameters in Table~\ref{tab:input} there
is a good agreement with the very accurate data \cite{Amhis:2012bh}:
\be\label{exp}
\Delta M_s = 17.69(8)\,\text{ps}^{-1}, \qquad \Delta M_d = 0.510(4)\,\text{ps}^{-1}~,
\ee
even if both central values are by $5-6\%$ above the data.
With the most recent values from the Twisted Mass Collaboration \cite{Carrasco:2013zta}
\be\label{twist}
\sqrt{\hat B_{B_s}}F_{B_s} = 262(10)\mev,\qquad \sqrt{\hat B_{B_d}}F_{B_d} = 216(10)\mev,
\ee
that are not yet included in the FLAG average, the central value of
$\Delta M_s$ would go down to $18.2/{\rm ps}$.

Concerning $S_{\psi\phi}$ and $S_{\psi K_S}$ we have
\be
S_{\psi\phi}= -\left(0.04^{+0.10}_{-0.13}\right), \qquad S_{\psi K_S}= 0.679(20)
\ee
with the second value known already for some time \cite{Amhis:2012bh} and the
first one being the most recent average from HFAG \cite{Amhis:2012bh} close
to the earlier result from the LHCb \cite{Aaij:2013oba}.
The first value is consistent with the SM expectation of $0.04$.  This is also
the case of  $S_{\psi K_S}$ for the values of $\vub$ and $\vcb$ used by us.

\boldmath
\subsection{$b\to s \mu^+\mu^-$ observables}
\unboldmath
The two Wilson coefficients that receive NP contributions in 331 models are
$C_9$ and $C_{10}$. We decompose them into the SM and NP contributions\footnote{These coefficients are defined as in
\cite{Buras:2012jb} and the same
definitions are used in  \cite{Descotes-Genon:2013wba,Altmannshofer:2013foa}.}:
\be
C_9=C_9^{\rm SM}+C_9^{\rm NP},\qquad C_{10}=C_{10}^{\rm SM}+C_{10}^{\rm NP},
\ee
where NP contributions have been given in (\ref{C9}) and (\ref{C10}) and
 the SM contributions are given as follows
\begin{align}
 \sin^2\theta_W C^{\rm SM}_9 &=\sin^2\theta_W P_0^{\rm NDR}+ [\eta_{\rm eff} Y_0(x_t)-4\sin^2\theta_W Z_0(x_t)],\\
   \sin^2\theta_W C^{\rm SM}_{10} &= -\eta_{\rm eff} Y_0(x_t) \label{Yeff}
 \end{align}
with all the entries given in \cite{Buras:2012jb,Buras:2013qja} except
for $\eta_{\rm eff}$ which is new and given below.
We have then
\be\label{CSM910}
C^{\rm SM}_9\approx 4.1, \qquad C^{\rm SM}_{10}\approx -4.1~.
\ee

In the case of $B_{s}\to\mu^+\mu^-$ decay one has  \cite{deBruyn:2012wk,Fleischer:2012fy,Buras:2013uqa}
\be\label{Rdef}
\frac{\overline{\mathcal{B}}(B_{s}\to\mu^+\mu^-)}{\overline{\mathcal{B}}(B_{s}\to\mu^+\mu^-)_{\rm SM}}
= \left[\frac{1+{\cal A}^{\mu\mu}_{\Delta\Gamma}\,y_s}{1+y_s} \right] |P|^2,
\qquad
P=\frac{C_{10}}{C_{10}^{\rm SM}}
\equiv |P|e^{i\varphi_P},
\ee
where 
\begin{align}
 &\overline{\mathcal{B}}(B_s\to\mu^+\mu^-)_\text{SM} = \frac{1}{1-y_s}\mathcal{B}(B_s\to\mu^+\mu^-)_\text{SM}\,,\\
&\mathcal{B}(B_s\to\mu^+\mu^-)_\text{SM} = \tau_{B_s}\frac{G_F^2}{\pi}\left(\frac{\alpha}{4\pi \sin^2\theta_W}\right)^2F_{B_s}^2m_\mu^2 
m_{B_s}\sqrt{1-\frac{4m_\mu^2}{m_{B_s}^2}}\left|V_{tb}^*V_{ts}\right|^2\eta_\text{eff}^2Y_0(x_t)^2\,,\\
&{\cal A}^{\mu\mu}_{\Delta\Gamma}=\cos(2\varphi_P-2\varphi_{B_s}),\qquad
	y_s\equiv\tau_{B_s}\frac{\Delta\Gamma_s}{2}
=0.062\pm0.009
\end{align}
with the later value being the latest world average \cite{Amhis:2012bh}.
The {\it bar} indicates that $\Delta\Gamma_s$ effects have been taken into account. In the SM and CMFV ${\cal
A}^{\mu\mu}_{\Delta\Gamma}=1$
but in the 331 models it is
slightly smaller and we take this effect into account. Generally as shown
in \cite{Buras:2013uqa} ${\cal A}^{\mu\mu}_{\Delta\Gamma}$ can serve to test
NP models as it can be determined in time-dependent
measurements \cite{deBruyn:2012wk,Fleischer:2012fy}.
Of interest is also the CP asymmetry
\begin{equation}\label{defys}
S^s_{\mu\mu}=\sin(2\varphi_P-2\varphi_{B_s}),
\end{equation}
which has been studied in detail in \cite{Buras:2012jb,Buras:2013uqa}
in the context of $Z^\prime$ models.
 In the case of $B_d\to\mu^+\mu^-$ decay the formulae given above apply with
$s$ replaced by $d$ and $y_d\approx 0$. Explicit formulae for $B_d\to\mu^+\mu^-$ can be found in
\cite{Buras:2012jb}.

Concerning the the status
 of the branching ratios for
$B_{s,d}\to \mu^+\mu^-$ decays we have
\be\label{LHCb2}
\overline{\mathcal{B}}(B_{s}\to\mu^+\mu^-)_{\rm SM}= (3.65\pm0.23)\cdot 10^{-9},
\quad
\overline{\mathcal{B}}(B_{s}\to\mu^+\mu^-) = (2.9\pm0.7) \times 10^{-9},
\ee
\be\label{LHCb3}
\mathcal{B}(B_{d}\to\mu^+\mu^-)_{\rm SM}=(1.06\pm0.09)\times 10^{-10}, \quad
\mathcal{B}(B_{d}\to\mu^+\mu^-) =\left(3.6^{+1.6}_{-1.4}\right)\times 10^{-10}.
\ee
The SM values are based on \cite{Bobeth:2013uxa} in which NLO corrections of electroweak
origin \cite{Bobeth:2013tba} and QCD corrections up to NNLO \cite{Hermann:2013kca} have been taken
into account. These values are rather close to the ones presented previously
by us \cite{Buras:2012ru,Buras:2013uqa} but the inclusion of these new
higher order corrections that were missing until now reduced significantly
various scale uncertainties so that non-parametric uncertainties in both branching
ratios are below $2\%$. The  experimental data are the most recent averages of the results from LHCb and CMS
\cite{Aaij:2013aka,Chatrchyan:2013bka,CMS-PAS-BPH-13-007}.

  The calculations performed in \cite{Bobeth:2013tba,Hermann:2013kca}
are very involved and in analogy to the QCD factors, like $\eta_B$ and $\eta_{1-3}$ in $\Delta F=2$ processes, we find it 
 useful to include all QCD and electroweak 
corrections into  $\eta_{\rm eff}$ introduced in (\ref{Yeff}) that without 
these corrections would be equal to unity.
Inspecting the analytic formulae in \cite{Bobeth:2013uxa} one finds then
\be\label{etaeff}
\eta_{\rm eff}= 0.9882\pm 0.0024~.
\ee

The small departure of  $\eta_{\rm eff}$ from unity  was already
anticipated in \cite{Misiak:2011bf,Buras:2012ru} but only the calculations in
\cite{Bobeth:2013uxa,Bobeth:2013tba,Hermann:2013kca} could put these expectations and conjectures on firm footing.
Indeed, in order to end up with such a simple result it was crucial to
perform such  involved calculations as these small corrections are only
valid for particular definitions of the top-quark mass and of other electroweak
parameters involved. In particular one has to use in 
$Y_0(x_t)$ the $\overline{\rm MS}$-renormalized top-quark mass $m_t(m_t)$ with respect to QCD but on-shell with respect to electroweak interactions. This means
$m_t(m_t)=163.5\gev$  as calculated in  \cite{Bobeth:2013uxa}. Moreover, in 
using (\ref{etaeff}) to calculate observables like branching ratios it is 
important to have the same normalization of effective Hamiltonian as in 
the latter paper. There this normalization is expressed in terms of $G_F$ 
and $M_W$ only.  Needless to say one can also use directly the formulae in  \cite{Bobeth:2013uxa}. 

In the present paper we follow the normalization of effective Hamiltonian in 
\cite{Buras:1998raa} which uses $G_F$, $\alpha(M_Z)$ and $\sin^2\theta_W$ 
and in order to be consistent with the calculation in \cite{Bobeth:2013uxa} 
our $\eta_{\rm eff}= 0.991$ with $m_t(m_t)$ unchanged. Interestingly also in the case of $\kpn$ and $\klpn$ the analog of $\eta_{\rm eff}$, multiplying this time $X_0(x_t)$, is found  with the  normalizations of effective Hamiltonian in \cite{Buras:1998raa}  and definition of $m_t$ as given above to be within $1\%$ from unity \cite{Brod:2010hi}.

In the case of $B\to K^*\mu^+\mu^-$ we will concentrate our discussion on
the Wilson coefficient $C_9^{\rm NP}$ which can be extracted from
the angular observables, in particular $\langle F_L\rangle$, $\langle S_5\rangle$ and $\langle A_8 \rangle$,
in which within the 331 models  NP contributions enter
exclusively through this coefficient. On the other hand $\IM(C_{10}^{\rm NP})$
governs the CP-asymmetry $\langle A_7 \rangle$. Useful approximate
expressions for these  angular observables at low $q^2$ in terms of  $C_9^{\rm NP}$,
$C_{10}^{\rm NP}$ and other Wilson coefficients  have been provided in \cite{Altmannshofer:2013foa}. Specified to 331 models they
are given as follows
\begin{align}
&\langle F_L \rangle \approx 0.77 +0.05~ \RE\,C_9^{\rm NP},\\
&
\langle S_4 \rangle \approx 0.29 ,\\
&\langle S_5 \rangle \approx -0.14 -0.09~ \RE\,C_9^{\rm NP}.
\end{align}

\begin{align}
&\langle A_7 \rangle \approx  0.07~\IM\,C_{10}^{\rm NP},\\
&
\langle A_8 \rangle \approx 0.04~ \IM\,C_9^{\rm NP},\\
&\langle A_9 \rangle \approx 0.
\end{align}

Note that NP contributions to $\langle S_4 \rangle$ and  $\langle A_9 \rangle$
vanish in 331 models due to the absence of right-handed currents in these
models.

 Eliminating $\RE\,C_9^{\rm NP}$ from these expressions in favour of $\langle S_5\rangle$ one finds \cite{Buras:2013qja}
\be\label{bsmumuc}
\langle F_L \rangle=0.69-0.56 \langle S_5 \rangle,
\ee
which shows analytically the point made
 in \cite{Descotes-Genon:2013wba,Altmannshofer:2013foa} that
NP effects in $F_L$ and $S_5$ are anti-correlated as observed in the data.

Indeed, the recent $B\to K^*\mu^+\mu^-$ anomalies imply the following
ranges for  $C_9^{\rm NP}$  \cite{Descotes-Genon:2013wba,Altmannshofer:2013foa} respectively
\be\label{ANOM}
 C_9^{\rm NP}=-(1.6\pm0.3), \qquad
 C_9^{\rm NP}=-(0.8\pm0.3)~.
\ee
As
 $C_9^{\rm SM}\approx 4.1$ at $\mu_b=4.8\gev$, these are very significant
suppressions of this coefficient.  We note that $C_9$ remains real as in the
SM but the data do not yet preclude a significant imaginary part for this
coefficient. The details behind these two results that
differ by a factor of two is discussed in \cite{Altmannshofer:2013foa}.
In fact inspecting Figs.~3 and 4 of the latter paper one sees that if
the constraints from $A_{\rm FB}$ and $B\to K\mu^+\mu^-$ were not taken into account $C_9^{\rm NP}\approx -1.4$ alone could
explain the anomalies in
the observables $F_L$ and $S_5$. But the inclusion of these constraints
 reduces the size of this coefficient. Yet values of
$C_9^{\rm NP}\approx -(1.2-1.0)$ seem to give reasonable agreement with
all data and the slight reduction of departure of $F_L$ and $S_5$ from
their SM values in the future data would allow to explain the two anomalies
with the help of  $C_9^{\rm NP}$  only as suggested originally in
 \cite{Descotes-Genon:2013wba}.

 Similarly
a very recent comprehensive Bayesian analysis of
the authors of \cite{Beaujean:2012uj,Bobeth:2012vn} in \cite{Beaujean:2013soa} finds that although
SM works well, if one wants to interpret the data in extensions of the SM
then NP scenarios with dominant NP effect in $C_9$ are favoured although
the inclusion of chirality-flipped operators in agreement with \cite{Altmannshofer:2013foa} would help to reproduce the data.
This is also confirmed
in \cite{Buras:2013qja,Horgan:2013pva}.
 However, as we remarked at the beginning of our paper, a very recent analysis in \cite{Descotes-Genon:2013zva} challenges
the solution with significant right-handed currents and we are looking forward
to the consensus on this point in the future.
References to earlier papers on  $B\to K^*\mu^+\mu^-$
 by all these authors can be found in  \cite{Descotes-Genon:2013wba,Altmannshofer:2013foa,Bobeth:2012vn} and \cite{Buras:2013ooa}.

\begin{table}[!tb]
\center{\begin{tabular}{|l|l|}
\hline
$G_F = 1.16637(1)\times 10^{-5}\gev^{-2}$\hfill\cite{Nakamura:2010zzi} 	&  $m_{B_d}=
m_{B^+}=5279.2(2)\mev$\hfill\cite{Beringer:1900zz}\\
$M_W = 80.385(15) \gev$\hfill\cite{Nakamura:2010zzi}  								&	$m_{B_s} =
5366.8(2)\mev$\hfill\cite{Beringer:1900zz}\\
$\sin^2\theta_W = 0.23116(13)$\hfill\cite{Nakamura:2010zzi} 				& 	$F_{B_d} =
(190.5\pm4.2)\mev$\hfill \cite{Colangelo:2010et}\\
$\alpha(M_Z) = 1/127.9$\hfill\cite{Nakamura:2010zzi}									&
$F_{B_s} =
(227.7\pm4.5)\mev$\hfill \cite{Colangelo:2010et}\\
$\alpha_s(M_Z)= 0.1184(7) $\hfill\cite{Nakamura:2010zzi}			&  $F_{B^+} =(185\pm3)\mev$\hfill
\cite{Dowdall:2013tga}
\\\cline{1-1}
$m_u(2\gev)=(2.1\pm0.1)\mev $ 	\hfill\cite{Laiho:2009eu}						&  $\hat B_{B_d}
=1.27(10)$,  $\hat
B_{B_s} =
1.33(6)$\hfill\cite{Colangelo:2010et}\\
$m_d(2\gev)=(4.73\pm0.12)\mev$	\hfill\cite{Laiho:2009eu}							& $\hat
B_{B_s}/\hat B_{B_d}
= 1.01(2)$ \hfill \cite{Carrasco:2013zta} \\
$m_s(2\gev)=(93.4\pm1.1) \mev$	\hfill\cite{Laiho:2009eu}				&
$F_{B_d} \sqrt{\hat
B_{B_d}} = 216(15)\mev$\hfill\cite{Colangelo:2010et} \\
$m_c(m_c) = (1.279\pm 0.013) \gev$ \hfill\cite{Chetyrkin:2009fv}					&
$F_{B_s} \sqrt{\hat B_{B_s}} =
266(18)\mev$\hfill\cite{Colangelo:2010et} \\
$m_b(m_b)=4.19^{+0.18}_{-0.06}\gev$\hfill\cite{Nakamura:2010zzi} 			& $\xi =
1.268(63)$\hfill\cite{Colangelo:2010et}
\\
$m_t(m_t) = 163.5(9)\gev$\hfill\cite{Beringer:1900zz} &  $\eta_B=0.55(1)$\hfill\cite{Buras:1990fn}  \\
$M_t=173.1(9) \gev$\hfill\cite{Aaltonen:2012ra}						&  $\Delta M_d = 0.510(4)
\,\text{ps}^{-1}$\hfill\cite{Amhis:2012bh}\\\cline{1-1}
$m_K= 497.614(24)\mev$	\hfill\cite{Nakamura:2010zzi}								&  $\Delta M_s =
17.69(8)
\,\text{ps}^{-1}$\hfill\cite{Amhis:2012bh}
\\
$F_K = 156.1(11)\mev$\hfill\cite{Laiho:2009eu}
	&
$S_{\psi K_S}= 0.679(20)$\hfill\cite{Nakamura:2010zzi}\\
$\hat B_K= 0.766(10)$\hfill\cite{Colangelo:2010et}
	&
$S_{\psi\phi}= -(0.04^{+0.10}_{-0.13})$\hfill\cite{Amhis:2012bh}\\
$\kappa_\epsilon=0.94(2)$\hfill\cite{Buras:2008nn,Buras:2010pza}				&
$\Delta\Gamma_s/\Gamma_s=0.123(17)$\hfill\cite{Amhis:2012bh}
\\
$\eta_{cc}=1.87(76)$\hfill\cite{Brod:2011ty}

	& $\tau_{B_s}= 1.516(11)\,\text{ps}$\hfill\cite{Amhis:2012bh}\\
$\eta_{tt}=0.5765(65)$\hfill\cite{Buras:1990fn}

& $\tau_{B_d}= 1.519(7) \,\text{ps}$\hfill\cite{Amhis:2012bh}\\
$\eta_{ct}= 0.496(47)$\hfill\cite{Brod:2010mj}

& $\tau_{B^\pm}= 1.641(8)\,\text{ps}$\hfill\cite{Amhis:2012bh}    \\
$\Delta M_K= 0.5292(9)\times 10^{-2} \,\text{ps}^{-1}$\hfill\cite{Nakamura:2010zzi}	&
$|V_{us}|=0.2252(9)$\hfill\cite{Amhis:2012bh}\\
$|\eps_K|= 2.228(11)\times 10^{-3}$\hfill\cite{Nakamura:2010zzi}					&
$|V_{cb}|=(42.4(9))\times
10^{-3}$\hfill\cite{Gambino:2013rza}\\
  $\mathcal{B}(B^+\to\tau^+\nu)=(0.96\pm0.26)\times10^{-4}$\hfill\cite{Amhis:2012bh}                                &
$|V_{ub}|=(3.6\pm0.3)\times10^{-3}$\hfill\cite{Beringer:1900zz}\\
\hline
\end{tabular}  }
\caption {\textit{Values of the experimental and theoretical
    quantities used as input parameters.}}
\label{tab:input}~\\[-2mm]\hrule
\end{table}

\section{Strategy for numerical analysis}\label{sec:4}
In our numerical analysis we will follow our recent strategy
applied to general LHS models in \cite{Buras:2013qja} with
the following significant simplification in the case of
331 models. The leptonic couplings of $Z^\prime$ are fixed
for a given $\beta$ and this allows us to avoid a rather involved
numerical analysis that  in  \cite{Buras:2013qja} had as a goal finding the optimal
values of these couplings. Even if $\beta$ is not fixed
and varying it changes the leptonic couplings in question,
the $\Delta_V^{\mu\bar\mu}$, $\Delta_A^{\mu\bar\mu}$ and
$\Delta_L^{\nu\bar\nu}$ couplings are correlated with each other
and finding one day their optimal values in 331 models will also
select the optimal value of $\beta$ fixing the electric
charges of new heavy gauge bosons and fermions. Of course
also quark couplings will play a prominent role in this
analysis, although even if they depend on $\beta$,
their correlation with leptonic couplings is washed out
by the new parameters in (\ref{newpar}).

Clearly NP contributions in any extension of the SM are constrained
by $\Delta F=2$ processes which presently are subject to theoretical
and experimental uncertainties. However, it
is to be expected that in the flavour precision era ahead of us, which will include both
advances in experiment and theory, in particular lattice calculations,
it will be possible to decide with high precision
whether $\Delta M_s$ and $\Delta M_d$ within the SM agree or disagree with
the data. For instance already the need for enhancements or suppressions
of these observables would be an important information. Similar comments
apply to $S_{\psi\phi}$ and $S_{\psi K_S}$ as well as to the branching
ratios $\mathcal{B}(B_{s,d}\to\mu^+\mu^-)$ and angular observables in
$B_d\to K^*\mu^+\mu^-$. In particular  correlations
and anti-correlations between  suppressions and enhancements allow
to distinguish between various NP models as can be illustrated with the DNA
charts proposed in  \cite{Buras:2013ooa}.

In order to monitor this progress in the context of the 331 models  we
will consider similarly to  \cite{Buras:2013qja} the following
five bins for $C_{B_s}$ and $C_{B_d}$ in (\ref{CBq})
\begin{align}\begin{split}\label{bins}
& C_{B_s}=0.90\pm0.01~\text{(yellow)}, \quad 0.96\pm 0.01~\text{(green)},\quad 1.00\pm 0.01~\text{(red)}, \\&
C_{B_s}= 1.04\pm 0.01~\text{(blue)}, \quad 1.10\pm 0.01~\text{(purple)}\end{split}
\end{align}
and similarly for $C_{B_d}$. This strategy avoids variations over non-perturbative parameters like $F_{B_s}\sqrt{\hat B_{B_s}}$
and can be executed here because
in  331 models these ratios have
a very simple form
\be
C_{B_s}=\frac{|S(B_s)|}{S_0(x_t)}, \qquad C_{B_d}=\frac{|S(B_d)|}{S_0(x_t)}
\ee
 and thanks to the presence of a single operator do not involve any non-perturbative uncertainties. Of course in order
to find out the experimental values of these ratios one has to handle these
uncertainties but this is precisely what we want to monitor in the coming
years.  The most recent update from Utfit collaboration reads
\be\label{UTfit}
C_{B_s}=1.08\pm 0.09, \qquad C_{B_d}=1.10\pm 0.17~.
\ee
However, it should be stressed that such values are sensitive
to the CKM input and in fact as seen in (\ref{DMS}) and (\ref{DMD}) with the
central
values of CKM parameters in Table~\ref{tab:input} we
would rather expect the central values of $C_{B_s}$ and $C_{B_d}$ to be below
unity. In order to have full picture we will not use the values in (\ref{UTfit})
but rather investigate how the results depend on the bins in (\ref{bins}).

Concerning $S_{\psi\phi}$ and $S_{\psi K_S}$ we will vary them in the ranges
\be\label{SpsiKs}
-0.14 \le S_{\psi \phi}\le 0.09,\qquad 0.639\le S_{\psi K_S}\le 0.719
\ee
corresponding to $1\sigma$ and $2\sigma$ ranges around the central experimental
values for $S_{\psi\phi}$ and $S_{\psi K_S}$, respectively.

Finally, in order to be sure that the lower bounds from LEP-II and LHC on $M_{Z^\prime}$ are satisfied we will present the results
for $\beta=\pm 1/\sqrt{3}$ and $\beta=\pm 2/\sqrt{3}$
for $M_{Z^\prime}=3\tev$. We will return to this issue in Section~\ref{sec:6}.
The scaling law in (\ref{scaling}) allows to translate
our results for observables in $B_s$ and $B_d$ decays into results for other choices of  $M_{Z^\prime}$.  As we have shown in
(\ref{master6})
and (\ref{master6a}) the $M_{Z^\prime}$
dependence cancels out in $\Delta X(K)$ and $\Delta Y(K)$.

\section{Numerical analysis}\label{sec:5}
\subsection{CMFV case}
It will be instructive to begin our numerical analysis with a particular
case, considered in \cite{Gauld:2013qja}, in which
\be
V_L=V_\text{CKM}.
\ee
In this case the CP-asymmetries $S_{\psi\phi}$ and $S_{\psi K_S}$ equal the SM
ones and the Wilson coefficients  $C_9^{\rm NP}$ and $C_{10}^{\rm NP}$ remain
real as in the SM. Moreover, having only two new variables to our disposal, $\beta$ and
$M_{Z^\prime}$, we find very concrete predictions and a number of correlations.

In presenting our results in this section we choose the following colour
coding for $\beta$:
\be\label{betacoding}
\beta= -\frac{2}{\sqrt{3}}~\text{(red)},\quad \beta= -\frac{1}{\sqrt{3}}~\text{(blue)},\quad \beta=
\frac{1}{\sqrt{3}}~\text{(green)}, \quad \beta= \frac{2}{\sqrt{3}}~\text{(yellow)}.
\ee
The cases of $\beta=\pm\sqrt{3}$ will be considered separately.

\begin{figure}[!tb]
 \centering
\includegraphics[width = 0.45\textwidth]{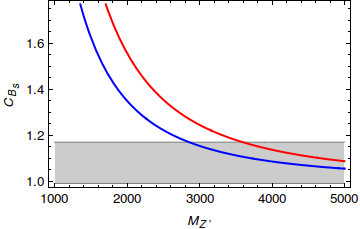}
\includegraphics[width = 0.45\textwidth]{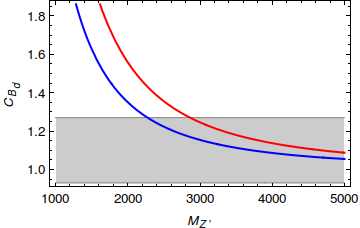}
\includegraphics[width = 0.45\textwidth]{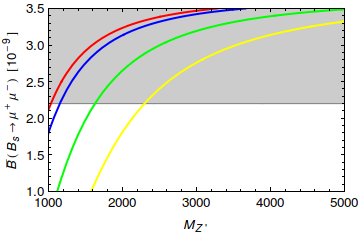}
\includegraphics[width = 0.45\textwidth]{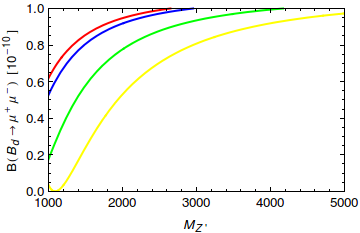}
\caption{$C_{B_s}$, $C_{B_d}$,  $\overline{\mathcal{B}}(B_s\to\mu^+\mu^-)$
and ${\mathcal{B}}(B_d\to\mu^+\mu^-)$ in the CMFV limit as functions of $M_{Z^\prime}$ for the four
chosen  values of $\beta$ with the colour coding given in (\ref{betacoding}). The gray regions show the UTfit 
range from Eq.~(\ref{UTfit}) and the experimental range $\overline{\mathcal{B}}(B_{s}\to\mu^+\mu^-) = (2.9\pm0.7)\cdot 10^{-9}$. 
The region for $B_d\to\mu^-\mu^-$ is outside the range of the plot.}\label{fig:F1}~\\[-2mm]\hrule
\end{figure}

\begin{figure}[!tb]
 \centering
\includegraphics[width = 0.45\textwidth]{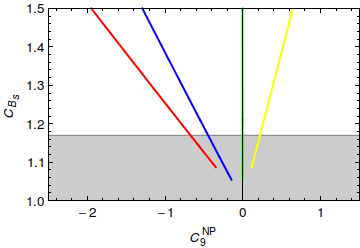}
\includegraphics[width = 0.45\textwidth]{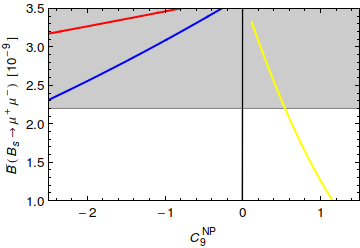}
\caption{Correlations
$C_{B_s}$ versus $C_9^{\rm NP}$ and
$\overline{\mathcal{B}}(B_s\to\mu^+\mu^-)$  versus $C_9^{\rm NP}$ in the CMFV limit
for  different values of $\beta$, varying $M_{Z^\prime}$ in the range
$2-5\,\tev$. Colour coding in (\ref{betacoding}). The gray regions show the UTfit 
range from Eq.~(\ref{UTfit}) and the experimental range $\overline{\mathcal{B}}(B_{s}\to\mu^+\mu^-) = (2.9\pm0.7)\cdot 10^{-9}$.
}\label{fig:F2}~\\[-2mm]\hrule
\end{figure}

In Fig.~\ref{fig:F1} we show  $C_{B_s}$, $C_{B_d}$,  $\overline{\mathcal{B}}(B_s\to\mu^+\mu^-)$
and ${\mathcal{B}}(B_d\to\mu^+\mu^-)$ as functions of $M_{Z^\prime}$ for the four
chosen  values of $\beta$. The values below $1.5\tev$ are presented only for illustration as such low masses appear rather
unrealistic on the basis of messages from the LHC.
In Fig.~\ref{fig:F2} we show the correlations
$C_{B_s}$ versus $C_9^{\rm NP}$ and
$\overline{\mathcal{B}}(B_s\to\mu^+\mu^-)$  versus $C_9^{\rm NP}$
for  different values of $\beta$, varying $M_{Z^\prime}$ in the range
$2-5\,\tev$.

\begin{figure}[!tb]
 \centering
\includegraphics[width = 0.45\textwidth]{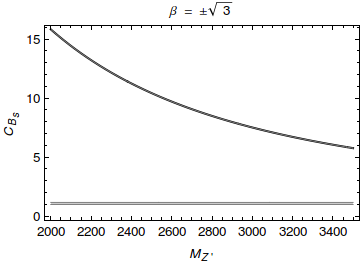}
\includegraphics[width = 0.45\textwidth]{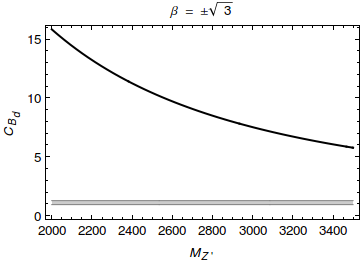}
\caption{$C_{B_s}$ and $C_{B_d}$ as functions of $M_{Z^\prime}$ for
$\beta=\pm\sqrt{3}$ in the CMFV limit. The gray regions show the UTfit 
range from Eq.~(\ref{UTfit}).}\label{fig:F8}~\\[-2mm]\hrule
\end{figure}

We observe:
\begin{itemize}
\item
 As already pointed out in \cite{Buras:2013qja} and known from CMFV scenario
 $C_{B_s}$ and  $C_{B_d}$ are bound to be above unity but this enhancement
for the values of $\beta$ in (\ref{betacoding})
 is not as severe as in the $\beta=-\sqrt{3}$ case considered in  \cite{Gauld:2013qja}.  It should be noted
that the sign of $\beta$ does not matter in these plots and the red and blue
lines shown there are equivalent to yellow and green lines, respectively.
\item
The case of $\beta=\pm \sqrt{3}$ is shown separately in
 Fig.~\ref{fig:F8} for fixed $s_W^2=0.246$, corresponding to $M_{Z^\prime}=2\tev$ only as an illustration.
Only values of $M_{Z^\prime}$ away from singularity are shown and to bring
$C_{B_s}$ and  $C_{B_d}$  down to the acceptable values  $M_{Z^\prime}$ has to
be increased well above  $M_{Z^\prime}=4\tev$ at which Landau singularity is present, that is beyond the range of validity of
this model.
The authors of \cite{Gauld:2013qja} working
with $s_W^2=0.231$ could not see these large enhancements of  $C_{B_s}$ and  $C_{B_d}$ . One can also check that for
$\beta=-\sqrt{3}$ and  $M_{Z^\prime}<3.5\tev$, in order to stay away from the singularity, $|C_9^{\rm NP}|$ is much larger than
indicated by the data. Clearly, as suggested in
\cite{Dias:2004wk,Dias:2004dc} one could improve this situation by shifting the
singularity above $4\tev$ through addition of other matter but then the model is a different one and one would have to
investigate what impact this additional matter has for observables considered here.
\item
As evident from our formulae for the $\Delta_A^{\mu\bar\mu}(Z^\prime)$ couplings
in the case of CMFV the branching ratios
$\overline{\mathcal{B}}(B_s\to\mu^+\mu^-)$  and ${\mathcal{B}}(B_d\to\mu^+\mu^-)$ can only be suppressed with respect to the SM.
This is welcome in the
case of $\overline{\mathcal{B}}(B_s\to\mu^+\mu^-)$ but not for ${\mathcal{B}}(B_d\to\mu^+\mu^-)$ where the present data would
favour an enhancement. But even
in the case  of $\overline{\mathcal{B}}(B_s\to\mu^+\mu^-)$  not all values of
$M_{Z^\prime}$ are consistent with the $1\sigma$ experimental range for
$\overline{\mathcal{B}}(B_s\to\mu^+\mu^-)$. In fact for the case $\beta=2/\sqrt{3}$ (yellow) values $M_{Z^\prime}\le 2\,\tev$ are
outside this range.
\item
The requirement of $C_9^{\rm NP}<0$ excludes in the CMFV case $\beta>0$.
The case of $\beta=-2/\sqrt{3}$ is clearly favoured as then the coupling
$\Delta_V^{\mu\bar\mu}(Z^\prime)$ is largest and values $C_{B_s}\approx 1.2$ would
be sufficient to soften significantly the $B_d\to K^*\mu^+\mu^-$ anomaly.
 For $\beta=-2/\sqrt{3}$ and  $M_{Z^\prime}\ge 2\,\tev$
 NP effects in $\overline{\mathcal{B}}(B_s\to\mu^+\mu^-)$  and ${\mathcal{B}}(B_d\to\mu^+\mu^-)$ are  small, in the ballpark of
$5-10\%$ of the SM values.
\end{itemize}

To summarize, in this scenario for quark couplings the case  $\beta=-2/\sqrt{3}$
is performing best as due to large value of $\Delta_V^{\mu\bar\mu}(Z^\prime)$ it
allows to obtain $C_9^{\rm NP}\approx -1.0$ for  $C_{B_s}\approx 1.2$.
 Yet a value  $C_{B_s}\approx 1.2$  could be problematic when the data improve. For $\beta=-1/\sqrt{3}$ which does not introduce
exotic charges the required values of $C_{B_s}$ to get sufficiently
negative  $C_9^{\rm NP}$ are even larger and the positive values of $\beta$
are excluded by the required sign of the latter coefficient.

Thus on the whole the idea of the authors of  \cite{Gauld:2013qja} to
consider negative values of $\beta$ was a good one but their choice
$\beta=-\sqrt{3}$ is excluded on the basis of the constraints on $C_{B_s}$ and $C_{B_d}$ when the correct values of
$\sin\theta_W^2$ at $M_{Z^\prime}$ are used. Moreover
LEP-II data on leptonic $Z^\prime$ couplings exclude this case as we will see
in the next section.

It should finally be noted that NP physics effects in Figs.~\ref{fig:F1} and
\ref{fig:F2} appear to
be significantly larger than found by us in \cite{Buras:2012dp}. One reason for
this
are different values of $\beta$ considered here but the primary reason is that the constraints
from $\Delta M_s$ and $\Delta M_d$ require the matrix $V_L$ to be even more
hierarchical than $V_{\rm CKM}$ and $V_L=V_{\rm CKM}$ in 331 models
appears to be problematic as we have just seen. The case  $V_L\not=V_{\rm CKM}$
is much more successful as we will demonstrate now.

\boldmath
\subsection{Non-CMFV case ($B_s$-system)}
\unboldmath
\subsubsection{Preliminaries}
Assuming next that the matrix $V_L$ in (\ref{VL-param}) differs from the CKM
matrix one has to find first the ranges of parameters (oases) for which
a given 331 model agrees with the $\Delta F=2$ data. The outcome of this search,
 for a given $C_{B_s}$ (or $C_{B_d}$ below),  are two oases in the space
$(\tilde s_{23}, \delta_2)$ for $B_s$-system and in the space $(\tilde s_{13}, \delta_1)$ for the $B_d$-system. We will not show
these oases as they have
structure similar to  the ones shown in \cite{Buras:2012dp,Buras:2012jb}. We only
recall that the two oases differ by a $180^\circ$ shift in the phases
$\delta_{1,2}$ which implies flips of signs of NP contributions to various $\Delta F=1$ observables. As the $\Delta F=2$
observables do not change under this
shift of phases, the correlation of a given $\Delta F=2$ observable like
$S_{\psi\phi}$ and  $\Delta F=1$ observable like  $\overline{\mathcal{B}}(B_s\to\mu^+\mu^-)$ in one oasis is changed to the
anti-correlation in another oasis and
vice versa. Measuring these two observables one can then determine the favoured
oasis and subsequently make predictions for other observables.

We will next
investigate what happens for the four different values of $\beta$ considered
by us and how the correlations between observables depend on the value
of $C_{B_s}$ using the colour coding in (\ref{bins}). We recall that
all results for $\beta=\pm 1/\sqrt{3}$ and  $\beta=\pm 2/\sqrt{3}$ are obtained for $M_{Z^\prime}=3\tev$ in order to be
sure that the LHC lower bound on $M_{Z^\prime}$ is satisfied, although as
we will discuss in the next section,  for  $\beta=\pm 1/\sqrt{3}$ also values $M_{Z^\prime}\approx 2.5\tev$ are consistent with these 
bounds.
 The results for
$\Delta F=1$ observables for
 $M_{Z^\prime}\not=3\tev$ can be obtained by using the scaling law in (\ref{scaling}). Then NP effects in $\beta=\pm
1/\sqrt{3}$ could still be
by a factor $1.2$ larger than shown in the plots below.

Concerning the case of  $\beta=-\sqrt{3}$, the problem with too a large
$C_{B_s}$ can now be avoided by properly choosing $V_L$ but the other
problems of this scenario, mentioned at the beginning of our paper and listed 
in Appendix~\ref{beta3}, cannot
be avoided in this manner and we will not discuss this case any more.

\boldmath
\subsubsection{Results}
\unboldmath

\begin{figure}[!tb]
 \centering
\includegraphics[width = 0.44\textwidth]{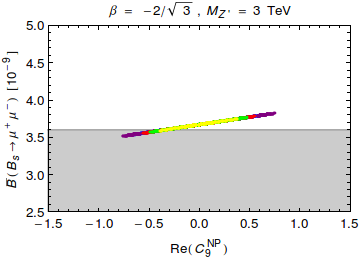}
\includegraphics[width = 0.44\textwidth]{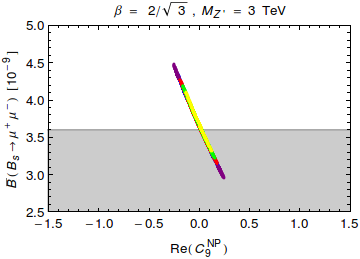}
\\
\includegraphics[width = 0.44\textwidth]{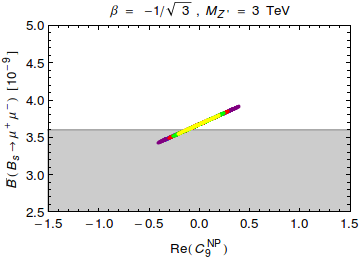}
\includegraphics[width = 0.44\textwidth]{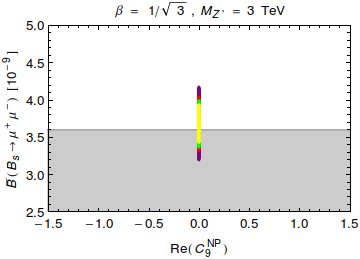}
\caption{Correlation
 $\overline{\mathcal{B}}(B_s\to\mu^+\mu^-)$ versus $\RE(C_9^\text{NP})$ for
$\beta=\pm 1/\sqrt{3}$ and $\beta=\pm 2/\sqrt{3}$ setting $M_{Z^\prime}=3\tev$ and different
values of $C_{B_s}$ with their colour coding in (\ref{bins}). The gray regions show  the experimental range  
$\overline{\mathcal{B}}(B_{s}\to\mu^+\mu^-) = (2.9\pm0.7)\cdot 10^{-9}$.
}\label{fig:F3}~\\[-2mm]\hrule
\end{figure}
\begin{figure}[!tb]
 \centering
\includegraphics[width = 0.46\textwidth]{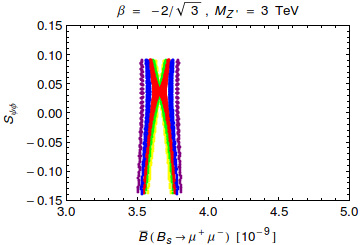}
\includegraphics[width = 0.46\textwidth]{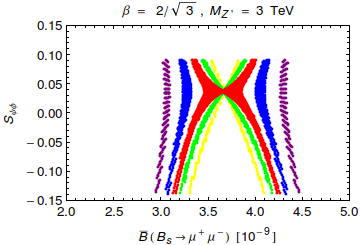}
\\
\includegraphics[width = 0.46\textwidth]{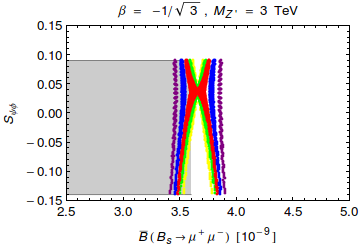}
\includegraphics[width = 0.46\textwidth]{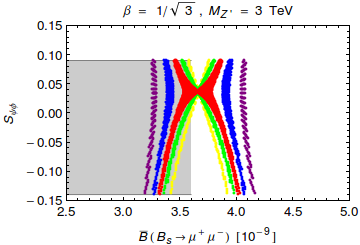}
\caption{Correlations
$S_{\psi\phi}$ versus $\overline{\mathcal{B}}(B_s\to\mu^+\mu^-)$ for
$\beta=\pm 1/\sqrt{3}$ and  $\beta=\pm 2/\sqrt{3}$ setting $M_{Z^\prime}=3\tev$ and different
values of $C_{B_s}$ with their colour coding in (\ref{bins}). The gray regions show  the experimental range for 
$\overline{\mathcal{B}}(B_{s}\to\mu^+\mu^-)$ and $S_{\psi\phi}$.
}\label{fig:F3a}~\\[-2mm]\hrule
\end{figure}
\begin{figure}[!tb]
 \centering
\includegraphics[width = 0.45\textwidth]{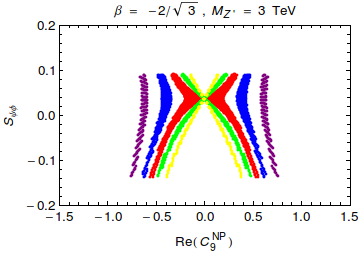}
\includegraphics[width = 0.45\textwidth]{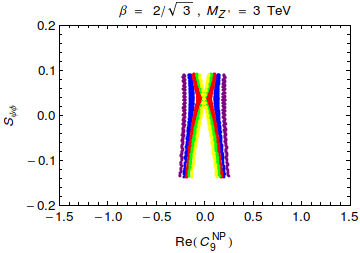}
\\
\includegraphics[width = 0.45\textwidth]{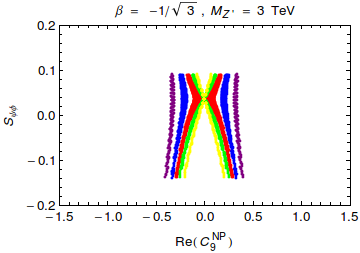}
\includegraphics[width = 0.45\textwidth]{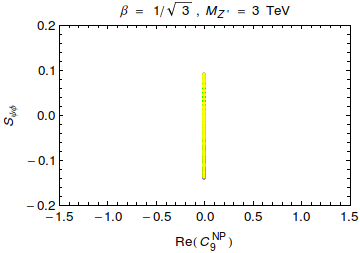}
\caption{Correlations
$S_{\psi\phi}$ versus $\RE(C_9^\text{NP})$
for $\beta=\pm 1/\sqrt{3}$ and
$\beta=\pm 2/\sqrt{3}$ setting $M_{Z^\prime}=3\tev$   and different
values of $C_{B_s}$ with their colour coding in (\ref{bins}).
}\label{fig:F3b}~\\[-2mm]\hrule
\end{figure}
\begin{figure}[!tb]
 \centering
\includegraphics[width = 0.45\textwidth]{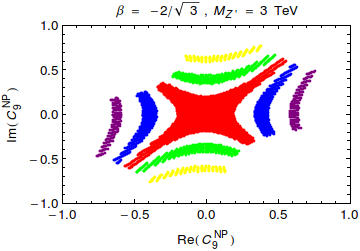}
\includegraphics[width = 0.45\textwidth]{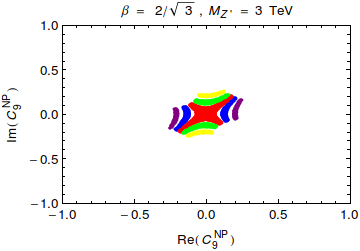}
\\
\includegraphics[width = 0.45\textwidth]{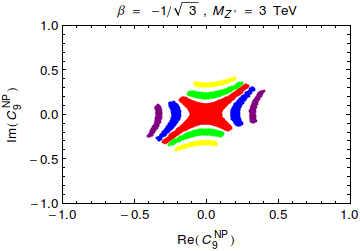}
\includegraphics[width = 0.45\textwidth]{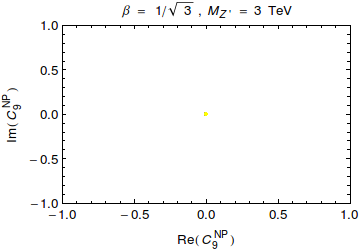}
\caption{Correlations
$\IM(C_9^\text{NP})$ versus $\RE(C_9^\text{NP})$
for $\beta=\pm 1/\sqrt{3}$ and
$\beta=\pm 2/\sqrt{3}$ setting $M_{Z^\prime}=3\tev$
and different
values of $C_{B_s}$ with their colour coding in (\ref{bins}).
}\label{fig:F3c}~\\[-2mm]\hrule
\end{figure}

\begin{figure}[!tb]
 \centering
\includegraphics[width = 0.45\textwidth]{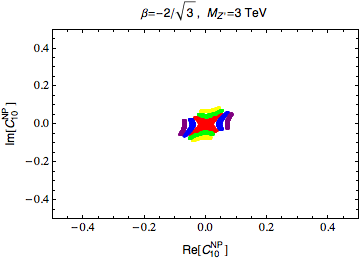}
\includegraphics[width = 0.45\textwidth]{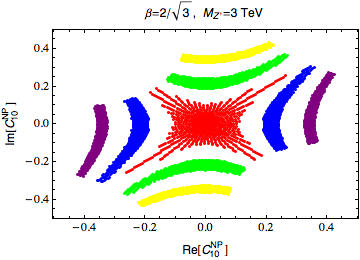}
\includegraphics[width = 0.45\textwidth]{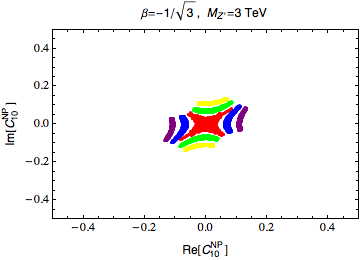}
\includegraphics[width = 0.45\textwidth]{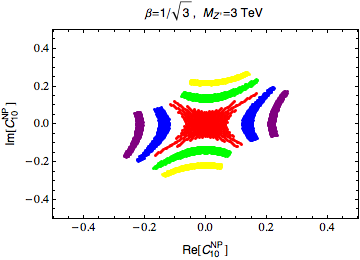}
\caption{ $\IM(C_{10}^{\rm NP})$ versus $\RE(C_{10}^{\rm NP})$
for $\beta=\pm 1/\sqrt{3}$ and
$\beta=\pm 2/\sqrt{3}$ setting $M_{Z^\prime}=3\tev$
and different
values of $C_{B_s}$ with their colour coding in (\ref{bins}).
}\label{fig:F11}~\\[-2mm]\hrule
\end{figure}

\begin{figure}[!tb]
 \centering
\includegraphics[width = 0.45\textwidth]{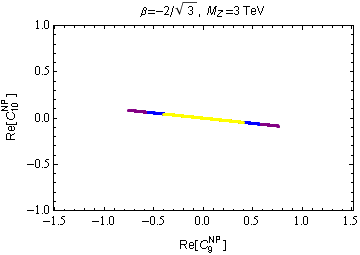}
\includegraphics[width = 0.45\textwidth]{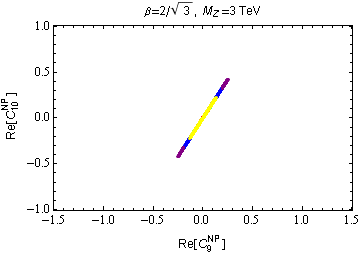}
\includegraphics[width = 0.45\textwidth]{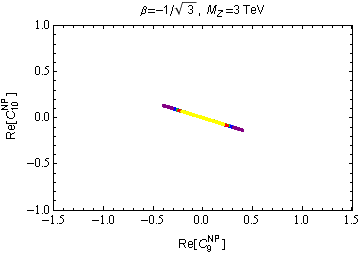}
\includegraphics[width = 0.45\textwidth]{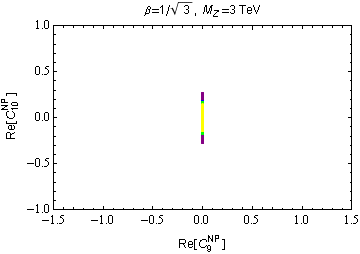}
\caption{ $\RE(C_{10}^{\rm NP})$ versus $\RE(C_{9}^{\rm NP})$
for $\beta=\pm 1/\sqrt{3}$ and
$\beta=\pm 2/\sqrt{3}$ setting $M_{Z^\prime}=3\tev$
and different
values of $C_{B_s}$ with their colour coding in (\ref{bins}).
}\label{fig:F12}~\\[-2mm]\hrule
\end{figure}

\begin{figure}[!tb]
 \centering
\includegraphics[width = 0.45\textwidth]{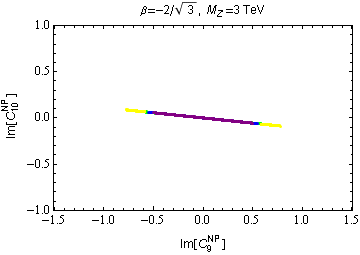}
\includegraphics[width = 0.45\textwidth]{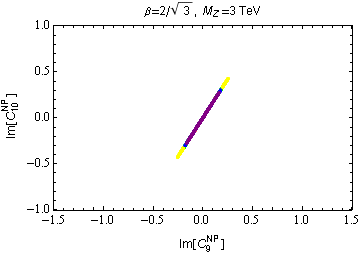}
\includegraphics[width = 0.45\textwidth]{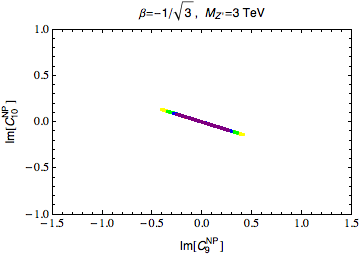}
\includegraphics[width = 0.45\textwidth]{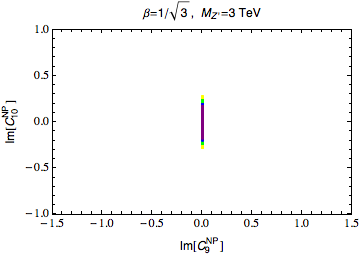}
\caption{ $\IM(C_{10}^{\rm NP})$ versus $\IM(C_{9}^{\rm NP})$
for $\beta=\pm 1/\sqrt{3}$ and
$\beta=\pm 2/\sqrt{3}$ setting $M_{Z^\prime}=3\tev$
and different
values of $C_{B_s}$ with their colour coding in (\ref{bins}).
}\label{fig:F13}~\\[-2mm]\hrule
\end{figure}

\begin{figure}[!tb]
 \centering
\includegraphics[width = 0.45\textwidth]{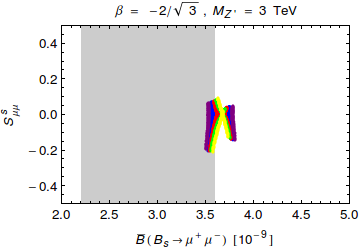}
\includegraphics[width = 0.45\textwidth]{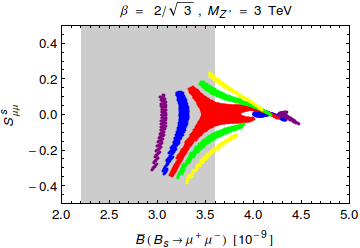}
\includegraphics[width = 0.45\textwidth]{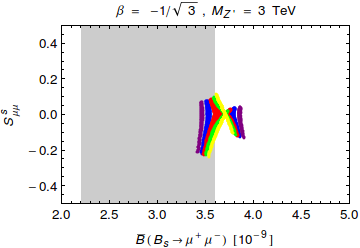}
\includegraphics[width = 0.45\textwidth]{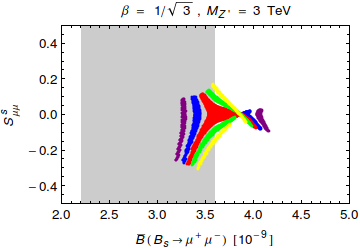}
\caption{ $S_{\mu\mu}^s$ versus $\overline{\mathcal{B}}(B_s\to\mu^+\mu^-)$
for $\beta=\pm 1/\sqrt{3}$ and
$\beta=\pm 2/\sqrt{3}$ setting $M_{Z^\prime}=3\tev$
and different
values of $C_{B_s}$ with their colour coding in (\ref{bins}). The gray regions show  the experimental range for 
$\overline{\mathcal{B}}(B_{s}\to\mu^+\mu^-)$.
}\label{fig:Smumu}~\\[-2mm]\hrule
\end{figure}

In Fig.~\ref{fig:F3} we show  $\overline{\mathcal{B}}(B_s\to\mu^+\mu^-)$ versus $\RE(C_9^\text{NP})$  for the four models considered.
These four plots exhibit
the structure identified through the ratio $R_1$ in (\ref{AV}) for which numerical values have been given in Table~\ref{tab:Ri}.
In particular we observe the
following features\footnote{We do not show the SM point in the plots as it
corresponds to the point where various curves cross each other and in any case
for coefficients that vanish in the SM it is obvious where the SM point is
placed in the plot.}:
\begin{itemize}
\item
For  a given $C_{B_s}\not=1$, one
can always find an oasis in which  $\RE(C_9^\text{NP})$ is negative
softening significantly the $B_d\to K^*\mu^+\mu^-$ anomalies. However
while for $\beta=-2/\sqrt{3}$ and $\beta=-1/\sqrt{3}$ the values
$\RE(C_9^\text{NP})=-0.8$ and $\RE(C_9^\text{NP})=-0.4$ can be reached respectively,
this is not possible for models with $\beta>0$.
\item
For $\beta <0$ the branching ratio
$\overline{\mathcal{B}}(B_s\to\mu^+\mu^-)$ remains SM-like although in
accordance with the relation (\ref{AV}) it is suppressed relative to its SM
value for negative $\RE(C_9^\text{NP})$. For the largest values of $C_{B_s}$ (purple and blue lines) this suppression can reach for most 
negative values of
$\RE(C_9^\text{NP})$  $4\%$ and $7\%$
for $\beta=-2/\sqrt{3}$ and $\beta=-1/\sqrt{3}$, respectively. The slope of the
strict correlation between these two observables depends on $\beta$. This
correlation is presently supported by the data for both observables even if
the effects in $\overline{\mathcal{B}}(B_s\to\mu^+\mu^-)$ are small.
\item
Looking at these four plots simultaneously we note that going from
 negative to positive values of $\beta$ the correlation line moves counter
clock-wise
with the center of the clock placed at the SM value. This of course means that
with increasing beta the correlation $\overline{\mathcal{B}}(B_s\to\mu^+\mu^-)$ versus $\RE(C_9^\text{NP})$ observed for $\beta
<0$ changes into anti-correlation for $\beta>0$, which is rather pronounced in the case of
$\beta=2/\sqrt{3}$. Consequently the suppression of $\overline{\mathcal{B}}(B_s\to\mu^+\mu^-)$ implies positive values of
$\RE(C_9^\text{NP})$ which is
not what we want to understand  $B_d\to K^*\mu^+\mu^-$ data. We also note
that for $\beta >0$ $\RE(C_9^\text{NP})$ remains
small but the effects in  $\overline{\mathcal{B}}(B_s\to\mu^+\mu^-)$ can
be larger than for $\beta <0$. These scenarios would be the favorite ones if the $B_d\to K^*\mu^+\mu^-$
anomalies decreased or disappeared  in the future while the experimental branching ratio
$\overline{\mathcal{B}}(B_s\to\mu^+\mu^-)$ turned out to be indeed by $20\%$
suppressed below its SM value as present central experimental and SM values seem
to indicate. In this case the model with $\beta=1/\sqrt{3}$ would be the
winner.
\end{itemize}

This pattern of effects for negative and positive values of $\beta$ is
also seen in Figs.~\ref{fig:F3a} and \ref{fig:F3b} where we show the
correlations between $S_{\psi\phi}$ versus $\overline{\mathcal{B}}(B_s\to\mu^+\mu^-)$ and $S_{\psi\phi}$ versus  $\RE(C_9^\text{NP})$,
respectively. In particular we
find that for models with $\beta <0$ for most negative values of  $\RE(C_9^\text{NP})$ and smallest values of
$\overline{\mathcal{B}}(B_s\to\mu^+\mu^-)$
 the negative values of  $S_{\psi\phi}$ are favoured. But as the values of $S_{\psi\phi}$ are
rather sensitive for a given value of $C_{B_s}$ to the value of $\overline{\mathcal{B}}(B_s\to\mu^+\mu^-)$, away from the lower
bound on this branching ratio
also positive values  of  $S_{\psi\phi}$ are allowed. This is in particular the case for largest values of $C_{B_s}$.

Bearing this ambiguity in mind, we identify therefore for a given $C_{B_s}$ a triple correlation between  $\RE(C_9^\text{NP})$,
$\overline{\mathcal{B}}(B_s\to\mu^+\mu^-)$ and  $S_{\psi\phi}$ that is an important test of this model. Interestingly the
requirement of a most negative $\RE(C_9^\text{NP})$ shifts automatically the other two observables closer to the data.

While the departure of $S_{\psi\phi}$ from its SM value is already a clear signal of new sources of CP-violation in $\Delta F=2$
transitions, non-vanishing
imaginary parts of $C_9$ and $C_{10}$ are signals of such new effects in
$\Delta F=1$ transitions. In Figs.~\ref{fig:F3c} and \ref{fig:F11} we
show the correlations $\IM(C_9^\text{NP})$ versus $\RE(C_9^\text{NP})$ and
$\IM(C_{10}^\text{NP})$ versus $\RE(C_{10}^\text{NP})$, respectively.  The fact that 
the pattern in both figures for a given $\beta$ is the same, even if the size of NP effects differs, is related to the relation 
(\ref{IMRE}).

We again observe that for $\beta<0$ NP effects are mainly seen in
$\IM(C_9^{\rm NP})$ while for $\beta>0$ in $\IM(C_{10}^{\rm NP})$.  In particular for $\beta=1/\sqrt{3}$ NP effects in $C_9$
practically vanish which is
a good test of this model.
Dependently
on the values of $C_{B_s}$ and $|\beta|$, the CP-asymmetry
$\langle A_8 \rangle $ could reach $(2-3)\%$  and the asymmetry
$\langle A_7 \rangle $ even $(3-4)\%$ for $\beta<0$ and $\beta>0$,
respectively.

 Finally in Figs.~\ref{fig:F12} and \ref{fig:F13} we show the correlations 
$\RE(C_{10}^{\rm NP})$ versus $\RE(C_9^{\rm NP})$ and 
$\IM(C_{10}^{\rm NP})$ versus $\IM(C_9^{\rm NP})$ for the four 331 models 
considered by us. These results follow from~(\ref{AV}).

New sources of CP-violation can also be tested in $B_s\to\mu^+\mu^-$
through the asymmetry $S_{\mu\mu}$ defined in (\ref{defys}) and
studied in detail in \cite{Buras:2012jb,Buras:2013uqa}
in the context of general $Z^\prime$ models. In Fig.~\ref{fig:Smumu} we
show the correlation of
$S_{\mu\mu}^s$ versus $\overline{\mathcal{B}}(B_s\to\mu^+\mu^-)$ in 331 model
considered. As expected the effects in the models with $\beta>0$ are
larger than for $\beta<0$. Similar to the case of $S_{\psi\phi}$ the required
suppression of $\overline{\mathcal{B}}(B_s\to\mu^+\mu^-)$ favours negative
values of $S^s_{\mu\mu}$ in all models.

As stressed in \cite{Altmannshofer:2013foa} the Wilson coefficient $C_9^{\rm NP}$ by itself has difficulty in removing completely the 
anomalies in 
$B_d\to K^*\mu^+\mu^-$ due to the constraint from $B_d\to K\mu^+\mu^-$. We have 
seen that even without this constraint the values of  $\RE~C_{9}^{\rm NP}$ 
have to be larger than $-0.8$ but this could turn out to be sufficient to 
reproduce the data when they improve. Still it is of interest to have a closer 
look at $B_d\to K\mu^+\mu^-$ within the four 331 models analysed by us.

To this end the approximate formula for the branching ratio confined to 
large $q^2$ region by  the authors of \cite{Altmannshofer:2013foa} is very useful. Lattice calculations of the relevant form factors 
are making significant progress here \cite{Bouchard:2013mia,Bouchard:2013eph} 
and the importance of this decay will increase in the future.
Neglecting the interference between NP 
contributions the formula of \cite{Altmannshofer:2013foa}  reduces in 
331 models to
\be\label{LHSK}
10^7\times \mathcal{B}(B_d\to K\mu^+\mu^-)_{[14.2,22]}=1.11+ 
0.27~\RE(C_9^\text{NP})\left(1-\frac{1}{R_1}\right),
\ee
where we have used (\ref{AV}) to express $\RE(C_{10}^\text{NP})$ in terms of 
$\RE(C_9^\text{NP})$.
The error on the first SM term is estimated to be $10\%$ \cite{Bouchard:2013mia,Bouchard:2013eph}. This should 
 be compared with the LHCb result
\be\label{equ:expBdKmu}
10^7\times \mathcal{B}(B_d\to K\mu^+\mu^-)_{[14.2,22]}=1.04\pm 0.12 \qquad {(\rm LHCb).}
\ee
 
\begin{figure}[!tb]
 \centering
\includegraphics[width = 0.44\textwidth]{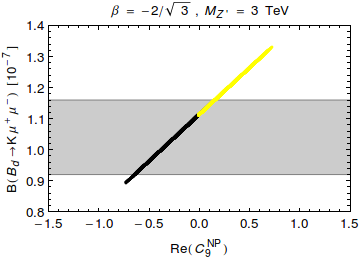}
\includegraphics[width = 0.44\textwidth]{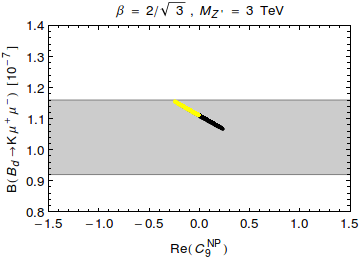}
\\
\includegraphics[width = 0.44\textwidth]{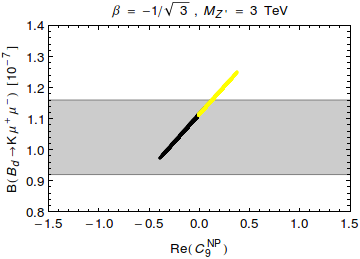}
\includegraphics[width = 0.44\textwidth]{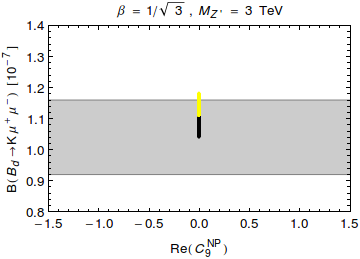}
\caption{Correlation
 $\mathcal{B}(B_d\to K\mu^+\mu^-)_{[14.2,22]}$  versus $\RE(C_9^{\rm NP})$ for
$\beta=\pm 1/\sqrt{3}$ and $\beta=\pm 2/\sqrt{3}$ setting $M_{Z^\prime}=3\tev$, 
$\overline{\mathcal{B}}(B_s\to\mu^+\mu^-)\leq\overline{\mathcal{B}}(B_s\to\mu^+\mu^-)_\text{SM}$ (black) and 
$\overline{\mathcal{B}}(B_s\to\mu^+\mu^-)\geq\overline{\mathcal{B}}(B_s\to\mu^+\mu^-)_\text{SM}$ (yellow) . The gray regions show  the 
experimental range for  $\mathcal{B}(B_d\to K\mu^+\mu^-)_{[14.2,22]}$ in~(\ref{equ:expBdKmu}).
}\label{fig:pBsmuvsBdKmu}~\\[-2mm]\hrule
\end{figure}

Using (\ref{LHSK}) we show
in Fig.~\ref{fig:pBsmuvsBdKmu}
  the correlation between $\mathcal{B}(B_d\to K\mu^+\mu^-)_{[14.2,22]}$ 
and $\RE(C_9^\text{NP})$ for the four 331 models in question. 

 We observe that the pattern of the correlations is similar to the ones 
in  Fig~\ref{fig:F3} which originates in the fact that $\mathcal{B}(B_d\to K\mu^+\mu^-)_{[14.2,22]}$ is strongly correlated with 
$\overline{\mathcal{B}}(B_s\to\mu^+\mu^-)$ within LHS models as already shown in \cite{Buras:2013qja} for a general LHS model. Moreover, as 
$R_1$ is fixed in a given model and its values have been collected in Table~\ref{tab:Ri} the straight lines in  Fig~\ref{fig:F3} can be 
easily understood.

There are two messages from this exercise: 
\begin{itemize}
\item
Our results for $\RE(C_9^{\rm NP})$ are fully in accordance with the present 
data on $\mathcal{B}(B_d\to K\mu^+\mu^-)_{[14.2,22]}$.
\item
On the basis of Figs.~\ref{fig:F3} and \ref{fig:pBsmuvsBdKmu} there is a 
triple correlation between  $\mathcal{B}(B_d\to K\mu^+\mu^-)_{[14.2,22]}$,  
$\RE(C_9^{\rm NP})$  and $\overline{\mathcal{B}}(B_s\to\mu^+\mu^-)$ which constitutes an important test for the models in question.  We 
indicate this 
correlation in Fig.~\ref{fig:pBsmuvsBdKmu} by showing when the latter 
branching ratio is suppressed ({\it black}) or enhanced ({\it yellow}) with respect to its SM value in 
accordance with the colour coding in DNA-charts of  \cite{Buras:2013ooa}.
\end{itemize}

\boldmath
\subsection{Non-CMFV case ($B_d$-system)}
\unboldmath
We have seen in the case of the MFV limit that ${\mathcal{B}}(B_d\to\mu^+\mu^-)$
is predicted to be suppressed relative to its SM value when  $\RE(C_9^\text{NP})$
is negative. This moves the theory away from the central value of the
experimental branching ratio. However, in the non-MFV case we can choose
the particular oasis in the space $(\tilde s_{13}, \delta_1)$ in which ${\mathcal{B}}(B_d\to\mu^+\mu^-)$  is enhanced.

\begin{figure}[!tb]
 \centering
\includegraphics[width = 0.45\textwidth]{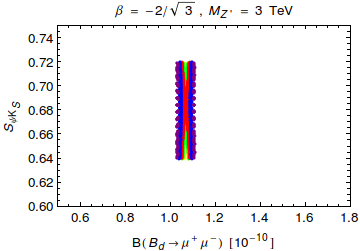}
\includegraphics[width = 0.45\textwidth]{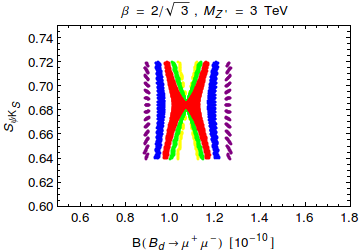}
\includegraphics[width = 0.45\textwidth]{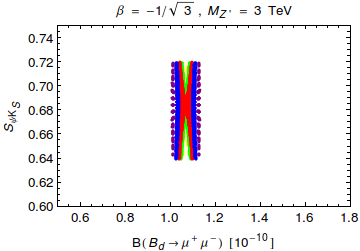}
\includegraphics[width = 0.45\textwidth]{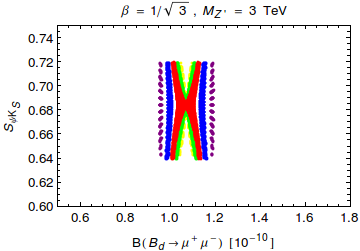}
\caption{Correlation
${\mathcal{B}}(B_d\to\mu^+\mu^-)$  versus $S_{\psi K_S}$ 
for $\beta=\pm 1/\sqrt{3}$ and
$\beta=\pm 2/\sqrt{3}$ setting $M_{Z^\prime}=3\tev$
and different
values of $C_{B_s}$ with their colour coding in (\ref{bins}).
}\label{fig:F5}~\\[-2mm]\hrule
\end{figure}

In the left  upper panel of Fig.~\ref{fig:F5} we show
${\mathcal{B}}(B_d\to\mu^+\mu^-)$  versus $S_{\psi K_S}$ again for $\beta=-2/\sqrt{3}$  and different bins of $C_{B_d}$. We observe
that the values of
${\mathcal{B}}(B_d\to\mu^+\mu^-)$ are SM-like and as already expected
from the values of $\Delta_A^{\mu\bar\mu}$ the central experimental
value of this branching ratio cannot be reproduced in this model.

More interesting results are found for $\beta > 0$. In
the right upper panel of Fig.~\ref{fig:F5} we show ${\mathcal{B}}(B_d\to\mu^+\mu^-)$  versus $S_{\psi K_S}$ for $\beta=2/\sqrt{3}$.
We observe that now enhancement of
${\mathcal{B}}(B_d\to\mu^+\mu^-)$ can reach $20\%$ over its SM value
but still far away from the central experimental value. For $\beta=-1/\sqrt{3}$
and $\beta=1/\sqrt{3}$ NP effects turn out to be larger and smaller relative
to $\beta=\mp 2/\sqrt{3}$ respectively, as one could deduce from the values
of the axial-vector couplings.

\begin{figure}[!tb]
 \centering
\includegraphics[width = 0.5\textwidth]{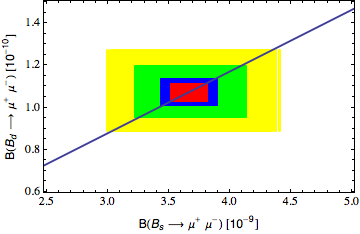}
\caption{Correlation
${\mathcal{B}}(B_d\to\mu^+\mu^-)$  versus
$\overline{\mathcal{B}}(B_s\to\mu^+\mu^-)$ for
the four models considered in the paper.  The colour coding
for $\beta$ is given
in (\ref{betacoding}). The straight line represents CMFV.
}\label{fig:F10}~\\[-2mm]\hrule
\end{figure}

Next in Fig.~\ref{fig:F10} we show
${\mathcal{B}}(B_d\to\mu^+\mu^-)$  versus $\overline{\mathcal{B}}(B_s\to\mu^+\mu^-)$ for  the four models considered with the colour coding
for $\beta$  given
in (\ref{betacoding}). We also show the CMFV line. As the uncertainty in the
latter line should be reduced to a few percent in this decade, this plot
could turn out to be useful for testing and distinguishing the four 331
models.

\boldmath
\subsection{Non-CMFV case for $b\to s\nu\bar\nu$, $\kpn$ and $\klpn$}
\unboldmath
\subsubsection{Preliminaries}
Finally, we turn our discussion to decays with neutrinos in the final state.
We recall that for  given $\beta$, $C_{B_d}$, $C_{B_s}$ and the chosen oases
in $B_d$ and $B_s$ systems the corresponding oasis including its size is fixed
so that definite predictions for $b\to s\nu\bar\nu$ transition, $\kpn$ and $\klpn$ can be made.

The inspection of the correlations presented in Section~\ref{sec:3a}
 teaches us about the following facts:
\begin{itemize}
\item
NP effects in $\varepsilon_K$ are  small but this is not a problem as
with our nominal values of $\vub$, $\vcb$ and $\gamma$ SM value of
$\varepsilon_K$ agrees well with the data.
\item
For $\beta>0$ NP effects in these decays are found to be small but are larger
in the cases with $\beta < 0$ where $Z^\prime$ couplings
to neutrinos are largest.
\item
Similarly NP effects  in $\kpn$ and $\klpn$ are small as we have already
expected on the basis  of the relation (\ref{master6}).
\end{itemize}

\boldmath
\subsubsection{The $b\to s\nu\bar\nu$ transitions}
\unboldmath

\begin{figure}[!tb]
 \centering
\includegraphics[width = 0.45\textwidth]{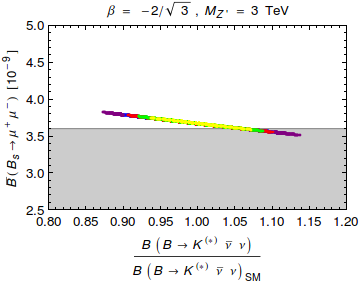}
\includegraphics[width = 0.45\textwidth]{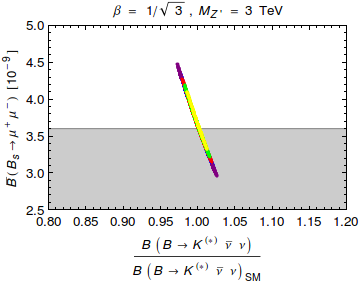}

\includegraphics[width = 0.45\textwidth]{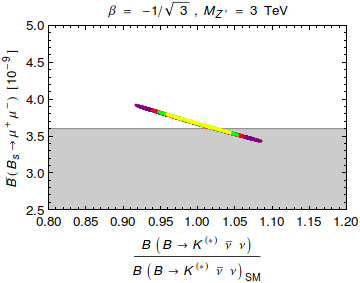}
\includegraphics[width = 0.45\textwidth]{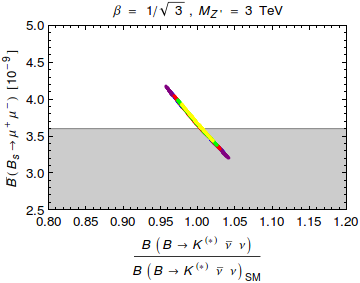}
\caption{ $\overline{\mathcal{B}}(B_s\to\mu^+\mu^-)$ versus the ratio in (\ref{Rnunu})   for all four $\beta=\pm\frac{2}{\sqrt{3}},
\pm
\frac{1}{\sqrt{3}}$.
}\label{fig:BsmuRnu}~\\[-2mm]\hrule
\end{figure}

In the absence of right-handed currents one finds \cite{Altmannshofer:2009ma}
\be\label{Rnunu}
R_{\nu\bar\nu}= \frac{\mathcal{B}(B\to K \nu \bar \nu)}{ \mathcal{B}(B\to K \nu \bar \nu)_{\rm SM}}=\frac{ \mathcal{B}(B\to K^* \nu \bar
\nu)}
 {\mathcal{B}(B\to K^* \nu \bar \nu)_{\rm SM}}=
 \frac{\mathcal{B}(B\to X_s \nu \bar \nu)}{\mathcal{B}(B\to X_s \nu \bar \nu)_{\rm SM}}=\frac{ |X_{\rm L}(B_s)|^2}{|\eta_X
X_0(x_t)|^2}~,
 \ee
with
\be
X_{\rm L}(B_s)=\eta_X X_0(x_t) +\Delta X(B_s)
\ee
and $\Delta X(B_s)$ given in (\ref{DXB}). The QCD factor $\eta_X=0.994$ 
\cite{Buchalla:1998ba}. { In this case the NLO electroweak corrections 
are of the order of one  per mil \cite{Brod:2010hi} when similarly to our discussion of $\eta_{\rm eff}$ in the context of $B_{s,d}\to\mu^+\mu^-$ decays one 
uses the normalization of effective Hamiltonian in \cite{Buras:1998raa} and
the top quark mass is evaluated in the $\overline{\rm MS}$ scheme for QCD 
and on-shell with respect to electroweak interactions. Thus accidentally 
$\eta_{\rm eff}$ that includes both QCD and electroweak corrections turns out 
in this scheme to be practically the same for $K\to\pi\nu\bar\nu$ and $B_{s,d}\to\mu^+\mu^-$ decays.}

The equality of these three ratios
is an important test of any LHS scenario. The violation of them would
imply the presence of right-handed couplings at work
\cite{Colangelo:1996ay,Buchalla:2000sk,Altmannshofer:2009ma}. In the context
of $Z^\prime$ models this is clearly seen in Fig.~20 of \cite{Buras:2012jb}.

The $SU(2)_L$ relation in (\ref{SU2}) satisfied in any LHS model, therefore  also in the 331 models presented by us, implies a
correlation between
$R_{\nu\bar\nu}$,  $\overline{\mathcal{B}}(B_s\to\mu^+\mu^-)$ and $C_9^{\rm NP}$
as shown for a general LHS model in Fig.~9 of \cite{Buras:2013qja}.

In Fig.~\ref{fig:BsmuRnu} we show one of these ratios versus $\overline{\mathcal{B}}(B_s\to\mu^+\mu^-)$ for  the models
considered. We observe that in
all models considered we have an
anti-correlation between these
two observables. But the predicted NP effects in all models
are rather small. The same conclusion has been reached for
general LHS models in
\cite{Altmannshofer:2013foa,Buras:2013qja}.

\begin{figure}[!tb]
 \centering
\includegraphics[width = 0.45\textwidth]{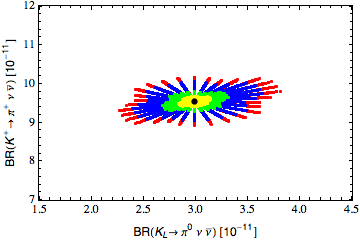}
\includegraphics[width = 0.45\textwidth]{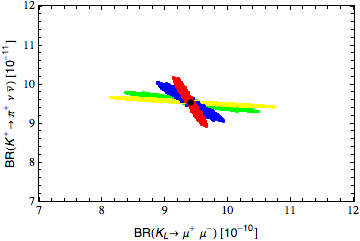}
\caption{{Correlations between rare $K$ decays for different values of
 $\beta$ using the colour coding in  (\ref{betacoding})}.
}\label{fig:Kpinu}~\\[-2mm]\hrule
\end{figure}

\boldmath
\subsubsection{$\kpn$, $\klpn$ and $K_L\to \mu^+\mu^-$ decays}
\unboldmath
The formulae for these decays have been given in \cite{Buras:2012jb} and will
not be repeated here.
In Fig.~\ref{fig:Kpinu} we show the correlation between
$\mathcal{B}(\kpn)$ and  $\mathcal{B}(\klpn)$ and the one
between
 $\mathcal{B}(\kpn)$ and  $\mathcal{B}(K_L\to\mu^+\mu^-)$ for the four models considered. The effects are rather small. What is
interesting are the SM values
in the middle of both plots that are enhanced over the usual values quoted
as a consequence of inclusive value of $\vcb$ used by us.

\section{Low and high energy constraints}\label{sec:6}
\subsection{Low energy precision observables}
Low energy precision observables provide additional bounds on the parameters of
 the models considered, in particular on the allowed range of $M_{Z^\prime}$ as
investigated  recently in the context of $\beta=-\sqrt{3}$ model in \cite{Gauld:2013qja}. We want to add that in concrete
models studied here the signs of deviations from SM predictions for these
observables are fixed providing additional tests beyond the lower bounds on
$M_{Z^\prime}$. In what follows we will present the predictions for three
such observables, considered also in \cite{Gauld:2013qja}, separately in each model from which the lower bounds on $M_{Z^\prime}$
follow.

We begin with the effect due to a $Z^\prime$ gauge boson on the weak charge of a nucleus
consisting of $Z$ protons and $N$ neutrons calculated in \cite{Bouchiat:2004sp}. In translating this result
into our notation one should note that the vector and axial-vector couplings
$f_{V,A}$ defined in \cite{Bouchiat:2004sp} are not equal to our couplings $\Delta_{V,A}(Z^\prime)$ but
are related through
\be
f_V=\frac{\Delta_{V}(Z^\prime)}{2}, \qquad f_A=-\frac{\Delta_{A}(Z^\prime)}{2}.
\ee
We find then ($\Delta_A^{e\bar e}(Z^\prime)=\Delta_A^{\mu\bar \mu}(Z^\prime)$)
\be
\Delta Q_W(Z,N)=\frac{1}{\sqrt{2}G_F}\frac{\Delta_A^{e\bar e}(Z^\prime)}{M_{Z^\prime}^2}
\left[(2Z+N)\Delta_V^{u\bar u}(Z^\prime)+(Z+2N)\Delta_V^{d\bar d}(Z^\prime)\right]
\ee
which  has an additional overall factor of $-1/4$ relative to
the corresponding expression in \cite{Gauld:2013qja} where $f_{V,A}=\Delta_{V,A}$  have been used\footnote{The authors of 
\cite{Gauld:2013qja} confirm our
 findings.}. We have then
\be
\Delta Q_W(Z,N)= (0.67)\, 10^{-2}\,\left[\frac{3\tev}{M_{Z^\prime}}\right]^2\Delta_A^{e\bar e}(Z^\prime)
\left[(2Z+N)\Delta_V^{u\bar u}(Z^\prime)+(Z+2N)\Delta_V^{d\bar d}(Z^\prime)\right].
\ee

Similarly for the effective shift in the weak charge of electron that can
be studied in M\o{}ller scattering we find
\be
\Delta Q^e_W= (0.67)\, 10^{-2}\,\left[\frac{3\tev}{M_{Z^\prime}}\right]^2\Delta_A^{e\bar e}(Z^\prime)
\Delta_V^{e\bar e}(Z^\prime).
\ee

For the violation of the first row CKM unitarity expressed through
\be
\tilde\Delta_{\rm CKM}\equiv 1-\sum_{q=d,s,b}|V_{uq}|^2
\ee
one has  for $M_{Z'}\gg M_W$ \cite{Marciano:1987ja,Gauld:2013qba,Buras:2013qja,Gauld:2013qja}
\be\label{Marciano}
\tilde\Delta_{\rm CKM}=\frac{3}{4\pi^2}\frac{M_W^2}{M_{Z'}^2}\ln\frac{M_{Z^\prime}^2}{M_W^2}
\Delta_L^{\mu\bar\mu}(Z^\prime)\left[\Delta_L^{\mu\bar\mu}(Z^\prime)-\Delta_L^{d\bar d}(Z^\prime)\right].
 \ee
In Table~\ref{tab:Zpred} we show predictions for these shifts in four
models considered by us and in each case the lower bound on $M_{Z^\prime}$
that follows from present experimental bounds.  In the first case
we use, as in  \cite{Gauld:2013qja}, Cesium nucleus with $Z=55$ and $N=78$.
We observe that the $90\%$ CL experimental bounds \cite{Beringer:1900zz,Marciano:1987ja,Davoudiasl:2012qa}
\be
|\Delta Q_W^{\rm Cs}|\le 0.6, \qquad |\Delta Q_W^{\rm e}|\le 0.016, \qquad
|\tilde\Delta_{\rm CKM}|\le 0.001
\ee
are well satisfied and the lower bounds on $M_{Z^\prime}$ are significantly
below the values used by us. We indicated
by dashes lower bounds on  $M_{Z^\prime}$ below $1\tev$.
In order to obtain these bounds we neglected
running of  $\sin^2\theta_W$ from $3\tev$ down to these bounds. Including it
would further weaken these bound but this effect is minor.

\begin{table}[!tb]
\centering
\begin{tabular}{|c||c|c|c|c|}
\hline
 $\beta$  & $1/\sqrt{3}$ & $-1/\sqrt{3}$ & $2/\sqrt{3}$ &  $-2/\sqrt{3}$ \\
\hline
\hline
  \parbox[0pt][1.6em][c]{0cm}{} $\Delta Q_W^{\rm Cs}$ & $0.106$ & $0.080$  & $0.158$ & $0.075$ \\
 \parbox[0pt][1.6em][c]{0cm}{} ${\rm Min}(M_{Z^\prime})[\tev] $         &  $1.26$ & $1.10$ & $1.54$ & $1.06$\\
\hline
 \parbox[0pt][1.6em][c]{0cm}{} $10^3\times\Delta Q_W^{e}$ & $-0.002$  & $-0.334$ & $0.656$ & $-0.402$\\
 \parbox[0pt][1.6em][c]{0cm}{} ${\rm Min}(M_{Z^\prime})[\tev] $   & $-$  & $-$ & $-$ & $-$\\
\hline
 \parbox[0pt][1.6em][c]{0cm}{} $ 10^4\times\tilde\Delta_{\rm CKM}$ & $0.154$ & $0.482$  & $0.088$ & $1.13$ \\
 \parbox[0pt][1.6em][c]{0cm}{} ${\rm Min}(M_{Z^\prime})[\tev] $    &  $-$ & $-$ & $-$ & $1.01$ \\
 \hline
\end{tabular}
\caption{Prediction for various observables for different $\beta$ setting
$M_{Z^\prime}=3\tev$.
Only  lower
bounds on  $M_{Z^\prime}$ above $1\tev$ resulting from present constraints on these
observables are shown.
}\label{tab:Zpred}~\\[-2mm]\hrule
\end{table}

\subsection{LEP-II constraints}
Recently the final analysis of LEP-II data by the LEP electroweak working group  appeared in \cite{Schael:2013ita} which allows
us to check whether the
values for $M_{Z^\prime}$ for the six 331 models considered by us are
consistent with these data. The data relevant for
us correspond to the range of center of mass energy $189\gev\le\sqrt{s}\le207\gev$. In our numerical calculations we will set
$\sqrt{s}=200\gev$.

The
fundamental for this analysis is the formula (3.8) in this paper \cite{Eichten:1983hw}\footnote{We prefer not to use $e^+ e^- \to
e^+e^-$ due to other
contributions like Babha scattering.}
\be
\mathcal{L}_{\rm eff}=\frac{4\pi}{\Lambda_\pm^2}\sum_{i,j=L,R}\eta_{ij}
\bar e_i\gamma_\mu e_i \bar f_j\gamma^\mu f_j, \qquad (f\not=e).
\ee
In this formula
$\eta_{ij}=\pm 1$ or $\eta_{ij}=0$. The different signs of $\eta_{ij}$ allow
to distinguish between constructive $(+)$ and destructive $(-)$ interference
between
the SM and NP contribution. $\Lambda_\pm$ is the scale of the contact interaction which can be related to $M_{Z^\prime}$ after
proper rescaling of $\eta_{ij}$. The lower bounds on $\Lambda_\pm$
 presented in table 3.15 of \cite{Schael:2013ita}  apply to certain choices of  $\eta_{ij}$ that are defined in table 3.14 of
that paper.

In the models considered by us there is the overall minus sign due
to $Z^\prime$ propagator  relative to the SM contribution which we include in
the definition of $\eta_{ij}$ so that with
\be\label{etaij}
\eta_{ij}^{ef}(Z^\prime)=-\Delta^{\e\bar\e}_i(Z^\prime)\Delta^{f\bar f}_j(Z^\prime),
\ee
 we obtain the relation
\be\label{ML}
M_{Z^\prime}=\frac{\Lambda_\pm}{\sqrt{4\pi}}\sqrt{|\Delta^{\e\bar\e}_i(Z^\prime)\Delta^{f\bar f}_j(Z^\prime)|}.
\ee
As we know the signs of $\eta_{ij}$ in each model we know in each case
whether the bound on $\Lambda_+$ or $\Lambda_-$ should be used. In Tables~\ref{tab:ll}--\ref{tab:uu} we list the values of the
couplings $\eta_{ij}$ for the six models considered by us together with the corresponding values for $\eta_{ij}(Z)$ for which the
minus sign in (\ref{etaij}) should be omitted as the energies
involved at LEP-II $\sqrt{s}>M_Z$.

The case of $\beta=-\sqrt{3}$ is easy to test in the case of $e^+e^-\to\mu^+\mu^-$ as in this case we deal with the model $VV^-$
of  \cite{Schael:2013ita}.
We find then the lower bound for $M_{Z^\prime}$ of $11\tev$, well above the
validity of this model. {  We would like to emphasize that this bound is quoted 
here only as an illustration. As discussed in Appendix~\ref{beta3} the coupling 
$\alpha_X$ at scales above $1\tev$ is too large to trust perturbation 
theory and calculating only tree diagrams misrepresents the real situation. 
Whether a non-perturbative dynamics would cure this model remains to be seen.}

Before turning to explicit four models analyzed by us let us note that in the
$LL^-$, $RR^-$ and $VV^-$ models for couplings in  \cite{Schael:2013ita},
which correspond to $\Delta_R^{l\bar l}(Z^\prime)=0$, $\Delta_L^{l\bar l}(Z^\prime)=0$ and $\Delta_L^{l\bar l}(Z^\prime)=\Delta_R^{l\bar
l}(Z^\prime)$ , respectively,  the combination of (\ref{ML}) and (\ref{SC9}) allows
to derive  upper bounds on $|C_9^{\rm NP}|$ that go beyond the 331 models and
apply  to LHS scenario for $Z^\prime$ generally. Indeed for the case
$e^+e^-\to\mu^+\mu^-$ we obtain from (\ref{ML}) the bound
\be\label{bound1}
\frac{M_{Z^\prime}}{\Delta_V^{\mu\bar\mu}}\ge  a\frac{\Lambda_-}{\sqrt{4\pi}}\equiv K
\ee
with a=1 for $LL^-$ and $RR^-$ and $a=1/2$ for $VV^-$. From table 3.15 in
 \cite{Schael:2013ita} we find  then
\be
K=2.77\tev~({\rm LL^-}),\qquad K=2.62\tev~({\rm RR^-}),\qquad  K=2.30\tev~({\rm VV^-}).
\ee
Therefore (\ref{SC9}) can be rewritten as an upper bound on $|C_9^{\rm NP}|$
as follows:
\be\label{C9bound}
|C_9^{\rm NP}|\le \frac{2.52\tev}{K}\sqrt{\frac{|\Delta S|}{0.231}}.
\ee
The last factor becomes unity for a $10\%$ contribution from NP to $\Delta M_s$
and consequently in this case the maximal by LEP-II allowed values for $|C_9^{\rm NP}|$ read: $0.91$, $0.96$ and $1.10$,  for  $LL^-$, 
$RR^-$
and $VV^-$, respectively. The latter case is the one considered in  \cite{Descotes-Genon:2013wba} and
also has similar structure to $\beta=-\sqrt{3}$ model without specification
of actual values of the muon couplings.

We conclude that for a $10\%$ shift in $S$ it is impossible in these models
to obtain $C_9^{\rm NP}=-1.5$ as found in  \cite{Descotes-Genon:2013wba}. Only
for effects $S$ in the ballpark of $20\%$ could such large negative values of
$C_9^{\rm NP}$ be obtained. While these results look similar to the ones
shown in Fig.~\ref{fig:F2}, they are more general as they do not assume CMFV and 331 models at work
and moreover take into account LEP-II data. Needless to say these LEP-II bounds can be significantly weakened by breaking
lepton universality in $Z^\prime$ couplings and suppressing $Z^\prime$
 couplings to electrons relative to the muon ones.

As far as the bound on $|C_{10}^{\rm NP}|$ is concerned the bounds obtained
for $LL^-$ and $RR^-$ apply also to this coefficient with $\Delta_V$ replaced by $\Delta_A$. For $VV^-$ this coefficient
vanishes. But for the case $AA^-$ in  \cite{Schael:2013ita} that corresponds
to  $\Delta_L^{l\bar l}(Z^\prime)=-\Delta_R^{l\bar l}(Z^\prime)$, we find
the analogue of the ratio $K$ to be $1.89\tev$ and slightly weaker bound than for
$|C_9^{\rm NP}|$ in the $VV^-$ case. Thus LEP-II bounds on  $|C_{10}^{\rm NP}|$
are weaker than the bounds presently available from $B_s\to\mu^+\mu^-$.

For the remaining models considered by us a complication arises due to the fact that the values of $\eta_{ij}$ in the simple
models studied in \cite{Schael:2013ita} and listed in table 3.14 of that
paper do not correspond to our models in which generally all combinations of
$L$ and $R$ contribute.

However, even without new global fits in these models, which would be beyond the scope of our paper we have checked by using the
Tables~\ref{tab:ll}--\ref{tab:uu}, the formulae in Appendix~\ref{app:LEP-II} and the table 3.15 of \cite{Schael:2013ita} that the
four models considered in detail by us satisfy all
LEP-II bounds. In fact our findings are as follows:
\begin{itemize}
\item
For the cases $n=-1,1,2$ the lower bounds on $M_{Z^\prime}$ are significantly
below $2\tev$, typically close to $1\tev$.
\item
For $\beta=-2\sqrt{3}$ the lower bound on $M_{Z^\prime}$ is below $2\tev$ but
its precise value would require a more sophisticated analysis. In any case
it appears that the LHC bound of approximately $3\tev$ in this model is stronger
that LEP-II bounds.
\end{itemize}

 Concerning the LHC bounds on  $M_{Z^\prime}$ from ATLAS and in particular
     CMS \cite{CMS-PAS-EXO-12-061}, the authors of \cite{Gauld:2013qja} using MAdGraph5 and CTEQ611 parton  distribution functions derived
for the
$\beta=-\sqrt{3}$ model a $95\%$ CL bound
 of $M_{Z^\prime}> 3.9\tev$. As these bounds are based on the Drell-Yan process
and are dominated by $Z^\prime$ couplings to up-quarks and muons that in 331
models equal to those of electrons, the values of $\eta_{ij}$ in Table~\ref{tab:uu} can give us a hint what happens in the models
considered by us.

As the relevant $\eta_{ij}$ in the models considered in detail by us are much
lower than the ones in the $\beta=-\sqrt{3}$ model, the lower bounds on $M_{Z^\prime}$ in these models
must be significantly lower than $3.9\tev$. On the other hand the couplings
in the $\beta=\pm 2/\sqrt{3}$ models are comparable, even if slightly larger
than the ones of $Z$ boson. Therefore, we expect that lower bound on $M_{Z^\prime}$
could be slightly larger than the one reported by CMS ($M_{Z^\prime}> 2.9\tev$)
and our choice of $M_{Z^\prime}=3.0\tev$ could be consistent with LHC bounds. Yet in order
to find it out a dedicated analysis would be necessary\footnote{Recently the bound $M_{Z^\prime}\ge 3.2\tev$ in this model resulting from the LHC has been 
derived 
 in \cite{Richard:2013xfa}.}. As far as
the LHC bounds for   $\beta=\pm 1/\sqrt{3}$ are concerned the analysis in
\cite{Coutinho:2013lta} indicates that in these models one could still have
$M_{Z^\prime}\approx 2.5\tev$. This would allow to enhance NP effects in
all $\Delta F=1$ observables in these models by roughly a factor of $1.2$
 {A complementary lower bound  $M_{Z^\prime}\ge 1\tev$  for 331 models with
$\beta = -1/\sqrt{3}$  was derived in \cite{Profumo:2013sca}
using dark matter data.}

As we have provided all information on the couplings necessary to
perform such an analysis in the $\beta=\pm 2/\sqrt{3}$ and $\beta=\pm 1/\sqrt{3}$ models, collider experimentalists and
phenomenologists  having the relevant codes could derive precise lower bounds on  $M_{Z^\prime}$ in the models in question. If
these bounds turn out in the future to be stronger or weaker than
$M_{Z^\prime}=3.0\tev$
our {\it scaling law} in (\ref{scaling}) will allow us to translate all results
presented in our paper into the new ones.

\begin{table}[!tb]
\centering
\begin{tabular}{|c||c|c|c|c|c|c|c|}
\hline
 $\beta$  & $1/\sqrt{3}$ & $-1/\sqrt{3}$ & $2/\sqrt{3}$ &  $-2/\sqrt{3}$ &
$\sqrt{3}$  &$-\sqrt{3}$& $Z$ couplings\\
\hline
\hline
  \parbox[0pt][1.6em][c]{0cm}{} LL  & $-0.168$ & $-0.666$  & $-0.068$ & $-1.65$ & $-0.01$ &$-62.0$& $0.396$\\
 \parbox[0pt][1.6em][c]{0cm}{} RR        &  $-0.165$ & $-0.165$ & $-1.05$ & $-1.05$ & $-0.03$ & $-60.5$&$0.296$\\
 \parbox[0pt][1.6em][c]{0cm}{} LR & $0.166$  & $-0.331$ & $0.267$ & $-1.32$ &$-0.02$ & $-61.2$& $-0.342$\\
 \parbox[0pt][1.6em][c]{0cm}{} RL & $0.166$  & $-0.331$ & $0.267$ & $-1.32$ &$-0.02$ & $-61.2$& $-0.342$\\
 \hline
\end{tabular}
\caption{Values of $10\times \eta_{ij}$  for different $\beta$ relevant
for $e^+e^-\to \ell^+\ell^-$ using
 $\sin^2\theta_W=0.249$ for
$\beta=\pm 1/\sqrt{3}$ and $\beta=\pm 2/\sqrt{3}$ and $\sin^2\theta_W=0.246$ for  $\beta=\sqrt{3}$. $\sin^2\theta_W=0.231$ for
$Z$-couplings.
}\label{tab:ll}~\\[-2mm]\hrule
\end{table}

\begin{table}[!tb]
\centering
\begin{tabular}{|c||c|c|c|c|c|c|c|}
\hline
 $\beta$  & $1/\sqrt{3}$ & $-1/\sqrt{3}$ & $2/\sqrt{3}$ &  $-2/\sqrt{3}$ &
$\sqrt{3}$  &$-\sqrt{3}$& $Z$ couplings\\
\hline
\hline
  \parbox[0pt][1.6em][c]{0cm}{} LL  & $0.223$ & $0.555$  & $0.157$ & $1.22$ & $0.23$ &$-41.6$& $0.626$\\
 \parbox[0pt][1.6em][c]{0cm}{} RR        &  $-0.055$ & $-0.055$ & $-0.351$ & $-0.351$ & $0.44$ & $-20.2$&$0.098$\\
 \parbox[0pt][1.6em][c]{0cm}{} LR & $0.055$  & $-0.110$ & $0.089$ & $-0.440$ &$0.22$ & $-20.5$&$-0.113$\\
 \parbox[0pt][1.6em][c]{0cm}{} RL & $-0.221$  & $0.276$ & $-0.618$ & $0.969$ &$0.45$ & $41.1$& $-0.542$\\
 \hline
\end{tabular}
\caption{Values of $10\times \eta_{ij}$  for different $\beta$ relevant
for $e^+e^-\to d\bar d$ using
 $\sin^2\theta_W=0.249$ for
$\beta=\pm 1/\sqrt{3}$ and $\beta=\pm 2/\sqrt{3}$ and $\sin^2\theta_W=0.246$ for  $\beta=\sqrt{3}$. $\sin^2\theta_W=0.231$ for
$Z$-couplings.
}\label{tab:dd}~\\[-2mm]\hrule
\end{table}

\begin{table}[!tb]
\centering
\begin{tabular}{|c||c|c|c|c|c|c|c|}
\hline
 $\beta$  & $1/\sqrt{3}$ & $-1/\sqrt{3}$ & $2/\sqrt{3}$ &  $-2/\sqrt{3}$ &
$\sqrt{3}$  &$-\sqrt{3}$& $Z$ couplings\\
\hline
\hline
  \parbox[0pt][1.6em][c]{0cm}{} LL  & $0.223$ & $0.555$  & $0.157$ & $1.22$ & $0.23$ &$-41.6$&$-0.511$\\
 \parbox[0pt][1.6em][c]{0cm}{} RR        &  $0.110$ & $0.110$ & $0.702$ & $0.702$ & $-0.88$ & $40.5$&$-0.198$\\
 \parbox[0pt][1.6em][c]{0cm}{} LR & $-0.111$  & $0.221$ & $-0.178$ & $0.86$ &$-0.44$ & $41.0$&$0.229$\\
 \parbox[0pt][1.6em][c]{0cm}{} RL & $-0.221$  & $0.276$ & $-0.618$ & $0.969$ &$0.45$ & $41.1$&$0.442$\\
 \hline
\end{tabular}
\caption{Values of $10\times \eta_{ij}$  for different $\beta$ relevant
for $e^+e^-\to u\bar u$ using
 $\sin^2\theta_W=0.249$ for
$\beta=\pm 1/\sqrt{3}$ and $\beta=\pm 2/\sqrt{3}$ and $\sin^2\theta_W=0.246$ for  $\beta=\sqrt{3}$. $\sin^2\theta_W=0.231$ for
$Z$-couplings.
}\label{tab:uu}~\\[-2mm]\hrule
\end{table}

\section{Summary and conclusions}\label{sec:7}
We have generalized our phenomenological
analysis of flavour observables in the particular
331 model with $\beta=1/\sqrt{3}$ presented in \cite{Buras:2012dp} to
the cases $\beta=-1/\sqrt{3}$, $\beta=\pm 2/\sqrt{3}$ and $\beta=\pm\sqrt{3}$
and confronted these models with the most recent data on $B_{s,d}\to \mu^+\mu^-$
and $B_d\to K^*\mu^+\mu^-$.  We have also presented predictions of these models for $b\to s \nu\bar\nu$ transitions and decays
$\kpn$, $\klpn$ and $K_L\to\mu^+\mu^-$.

Our three most important messages from this analysis are as follows:
\begin{itemize}
\item

The 331 models analyzed by us do not account for the $B_d\to K^*\mu^+\mu^-$ anomalies if the latter require $\RE(C_9^{\rm NP})\le -1.3$ as 
indicated by the model independent analysis in \cite{Descotes-Genon:2013wba}. On the other hand, these models could be in accordance with 
the outcome of the analyses in \cite{Altmannshofer:2013foa,Buras:2013qja,Horgan:2013pva,Beaujean:2013soa} provided the required size of 
$C_9^\prime$ in some of these papers will decrease with time (see also \cite{Descotes-Genon:2013zva} where the impact of a NP contribution 
to $C_9^\prime$ on these anomalies is discussed).
\item
Going beyond 331 models and assuming lepton
universality  we find an upper bound  $|C^{\rm NP}_{9}|\le 1.1 (1.4)$ from
LEP-II data for all $Z^\prime$ models within LHS scenario,  when NP contributions to $\Delta M_s$ at the level of  $10\%(15\%)$  are
allowed.
We conclude
therefore that it is unlikely that values like $\RE(C^{\rm NP}_{9})=-1.5$
can be accommodated in $Z^\prime$ models of LHS type when lepton universality
 is assumed. As the 331 models not analyzed by us belong to this class of
models, this finding applies to them as well.
\item
The central experimental value of ${\mathcal{B}}(B_d\to\mu^+\mu^-)$ from LHCb and CMS cannot be reproduced in the 331 models, although an 
enhancement by $20\%$
over its SM value is possible. A general LHS scenario can do much better
as demonstrated in \cite{Buras:2013qja}. But then the universality in
lepton couplings has to be broken to satisfy LEP-II constraints and the
diagonal $Z^\prime$ couplings to quarks must be smaller than in 331 models
considered by us to avoid the bounds on $M_{Z^\prime}$ from LHC.
\end{itemize}

In more detail our findings are as follows:
\begin{itemize}
\item
Analyzing the models with $\beta=\pm 1/\sqrt{3}$ and $\beta=\pm 2/\sqrt{3}$ we
 find that for $\beta>0$ measurable NP effects are allowed
 in $B_{s,d}\to \mu^+\mu^-$, sufficient to suppress
$\overline{\mathcal{B}}(B_{s}\to\mu^+\mu^-)$ down to its central experimental
value. On the other hand as mentioned above
${\mathcal{B}}(B_{d}\to\mu^+\mu^-)$ even if reaching
values $20\%$ above SM result, is still well below the experimental central
value.
Thus we expect that the experimental value of  ${\mathcal{B}}(B_{d}\to\mu^+\mu^-)$ must go down if these models should stay alive.
For $\beta>0$, the $B_d\to K^*\mu^+\mu^-$ anomaly cannot be explained and
in fact the anti-correlation between $\overline{\mathcal{B}}(B_{s}\to\mu^+\mu^-)$ and $\RE(C_9^{\rm NP})$ predicted in this case
is not in accordance with
the present data.  On the other hand in the case of
 the absence of $B_d\to K^*\mu^+\mu^-$ anomalies in the future
data and confirmation of the suppression of $\mathcal{B}(B_s\to\mu^+\mu^-)$
relative to its SM value the model with  $\beta=1/\sqrt{3}$ and $M_{Z^\prime}\approx 3\tev$  would be favoured.
\item
Presently, more interesting appear  models with $\beta<0$ where NP effects in $\overline{\mathcal{B}}(B_{s}\to\mu^+\mu^-)$ and
$\RE(C_9^{\rm NP})$ bring the theory closer  to the data. Moreover we identified  a triple correlation between  $\RE(C^{\rm
NP}_9)$, $\overline{\mathcal{B}}(B_{s}\to\mu^+\mu^-)$ and $S_{\psi\phi}$ that for
$\RE(C^{\rm NP}_{9})<-0.5$ required by $B_d\to K^*\mu^+\mu^-$ anomalies
 implies uniquely suppression of $\overline{\mathcal{B}}(B_{s}\to\mu^+\mu^-)$  relative to its SM value which is favoured by the
data. 
In turn also
$S_{\psi\phi}< S_{\psi\phi}^{\rm SM}$ is favoured with $S_{\psi\phi}$
having dominantly opposite sign to $S_{\psi\phi}^{\rm SM}$
 and closer to its central experimental value.  Figs.~\ref{fig:F3}-\ref{fig:F3b} show these correlations in explicit terms.
\item
Another important triple correlation is the one between $\RE(C^{\rm NP}_9)$, $\overline{\mathcal{B}}(B_{s}\to\mu^+\mu^-)$ and  $B_d\to K 
\mu^+\mu^-$.  It can be found in Fig.~\ref{fig:pBsmuvsBdKmu}.
\item
Our study of $b\to s\nu\bar\nu$ transitions,  $\kpn$ and $K_L\to\mu^+\mu^-$
shows that NP effects in these decays in the
models considered are typically below $10\%$ at the level of the
branching ratios. NP effects in $\klpn$ can reach $20\%$.
\item
We have demonstrated how the effects found by us are correlated with the departures of $C_{B_s}$ and  $C_{B_d}$ from unity. As
the latter departures depend sensitively on the precision of lattice non-perturbative calculations, the future
of 331 models does not only depend on experimental progress but also on
progress of latter calculations.
\item
As a by-product we have presented bounds on 331 models from low energy
precision experiments and provided enough information on the couplings of
$Z^\prime$ to quarks and leptons that a sophisticated analyses of  LEP-II
observables and of LHC constraints could be performed in the future.
\item
Finally, the model with $\beta=-\sqrt{3}$ can be ruled out on the basis of the data for
various
observables, in particular the final results from LEP-II. But even if
renormalization group effects in $\sin^2\theta_W$ are not taken into account,
the resulting lower bounds on $M_{Z^\prime}$ are  higher than the
upper bounds implied by the Landau singularity. On the other hand
the model with  $\beta=\sqrt{3}$ does not predict significant departures
from the SM.
\end{itemize}

Whether the models with $\beta=-1/\sqrt{3}$ and $\beta=-2/\sqrt{3}$ or
with $\beta=1/\sqrt{3}$ and $\beta=2/\sqrt{3}$ will be favoured by the data
will depend on the future of the experimental results for $B_{s,d}\to\mu^+\mu^-$, $B_d\to K^*(K)\mu^+\mu^-$ and  future values
of $C_{B_q}$. The numerous
plots presented in our paper should allow to monitor these developments.
Most importantly, the values of $M_{Z^\prime}$ considered in our paper are
sufficiently low that this new gauge boson could be discovered in the
next run of the LHC and its properties could even be studied at a future
ILC \cite{Richard:2013xfa}.

\section*{Acknowledgements}
We thank in particular Francois Richard for informative discussions on the bounds on $Z^\prime$  couplings from LEP-II data, 
Christoph Bobeth, Martin Gorbahn and Mikolaj Misiak for discussions
related to $B_s\to\mu^+\mu^-$ and David Straub in connection with the 
decay $B_d\to K^*(K)\mu^⁺\mu^⁻$. Extensive E-mail exchanges with Ulrich Haisch
in the context of the $\beta=-\sqrt{3}$ model were very enjoyable. Thanks go also to Quim Matias.
This research was done and financed in the context of the ERC Advanced Grant project ``FLAVOUR''(267104) and was partially
supported by the DFG cluster
of excellence ``Origin and Structure of the Universe''.

\appendix
\section{Expressions for couplings in various 331 models}\label{examples}
In obtaining the results below we use for $\beta=\pm 1/\sqrt{3}$ and
$\beta=\pm 2/\sqrt{3}$ the values $\sin^2\theta_W=0.249$ and $g=0.633$
corresponding to $M_{Z^\prime}=3\tev$. We stress that for these models
the dependence of the
couplings on $M_{Z^\prime}$ for  $1\tev\le M_{Z^\prime}\le 5\tev$, unless they are very small, is basically negligible assuring the scaling
law (\ref{scaling}).
\boldmath
\subsubsection*{$\beta=\pm 1/\sqrt{3}$}
\unboldmath
For both signs we have
\be
 \Delta_L^{ij}(Z') =\frac{g}{\sqrt{3}}c_W \sqrt{f(1/\sqrt{3})} v_{3i}^*v_{3j} =
 0.388~ v_{3i}^*v_{3j}
\ee

Next for $\beta=1/\sqrt{3}$ we have
{\allowdisplaybreaks
\begin{subequations}
\begin{align}
 \Delta_V^{d\bar d}(Z') &=\frac{g}{2\sqrt{3}c_W} \sqrt{f(1/\sqrt{3})}
~\left[-1+\frac{2}{3}s_W^2\right] = -0.215
 \,,\\
 \Delta_A^{d\bar d}(Z') &=\frac{g}{2\sqrt{3}c_W} \sqrt{f(1/\sqrt{3})}
~\left[1-2s_W^2\right]= 0.130 \,,\\
 \Delta_V^{u\bar u}(Z') &=\frac{g}{2\sqrt{3}c_W} \sqrt{f(1/\sqrt{3})}
~\left[-1+\frac{8}{3}s_W^2\right] = -0.087
 \,,\\
 \Delta_A^{u\bar u}(Z') &=\frac{g}{2\sqrt{3}c_W} \sqrt{f(1/\sqrt{3})}
~\left[1\right]= 0.258 \,,\\
\Delta_L^{\nu\bar\nu}(Z') & = \frac{g}{2\sqrt{3}c_W}\sqrt{f(1/\sqrt{3})} \left[1-2s_W^2\right]=  0.130 \,,\\
 \Delta_V^{\mu\bar\mu}(Z') &=\frac{g}{2\sqrt{3}c_W} \sqrt{f(1/\sqrt{3})}
~\left[1-4s_W^2\right] = 0.001
 \,,\\
 \Delta_A^{\mu\bar\mu}(Z') &=\frac{g}{2\sqrt{3}c_W} \sqrt{f(1/\sqrt{3})}
~\left[-1\right]= -0.258
\end{align}
\end{subequations}}%
 and for $\beta=-1/\sqrt{3}$
{\allowdisplaybreaks
\begin{subequations}
\begin{align}
\Delta_V^{d\bar d}(Z') &=\frac{g}{2\sqrt{3}c_W} \sqrt{f(1/\sqrt{3})}
~\left[-1+\frac{4}{3}s_W^2\right] = -0.172
 \,,\\
 \Delta_A^{d\bar d}(Z') &=\frac{g}{2\sqrt{3}c_W} \sqrt{f(1/\sqrt{3})}
~\left[1\right]= 0.258 \,,\\
 \Delta_V^{u\bar u}(Z') &=\frac{g}{2\sqrt{3}c_W} \sqrt{f(1/\sqrt{3})}
~\left[-1-\frac{2}{3}s_W^2\right] = -0.301
 \,,\\
 \Delta_A^{u\bar u}(Z') &=\frac{g}{2\sqrt{3}c_W} \sqrt{f(1/\sqrt{3})}
~\left[1-2s_W^2\right]= 0.130 \,,\\
\Delta_L^{\nu\bar\nu}(Z') & = \frac{g}{2\sqrt{3}c_W}\sqrt{f(1/\sqrt{3})} \left[1\right]= 0.258\,,\\
 \Delta_V^{\mu\bar\mu}(Z') &=\frac{g}{2\sqrt{3}c_W} \sqrt{f(1/\sqrt{3})}
~\left[1+2s_W^2\right] = 0.386
 \,,\\
 \Delta_A^{\mu\bar\mu}(Z') &=\frac{g}{2\sqrt{3}c_W} \sqrt{f(1/\sqrt{3})}
~\left[-1+2s_W^2\right]= -0.130
\end{align}
\end{subequations}}%

\boldmath
\subsubsection*{$\beta=\pm 2/\sqrt{3}$}
\unboldmath
For both signs we have
\be
 \Delta_L^{ij}(Z') =\frac{g}{\sqrt{3}}c_W \sqrt{f(2/\sqrt{3})} v_{3i}^*v_{3j} =
0.489~  v_{3i}^*v_{3j}
\ee

Next for $\beta=2/\sqrt{3}$ we have
{\allowdisplaybreaks
\begin{subequations}
\begin{align}
\Delta_V^{d\bar d}(Z') &=\frac{g}{2\sqrt{3}c_W} \sqrt{f(2/\sqrt{3})}
~\left[-1+\frac{1}{3}s_W^2\right] = -0.299
 \,,\\
 \Delta_A^{d\bar d}(Z') &=\frac{g}{2\sqrt{3}c_W} \sqrt{f(2/\sqrt{3})}
~\left[1-3s_W^2\right]= 0.082 \,,\\
 \Delta_V^{u\bar u}(Z') &=\frac{g}{2\sqrt{3}c_W} \sqrt{f(2/\sqrt{3})}
~\left[-1+\frac{13}{3}s_W^2\right] = 0.026
 \,,\\
 \Delta_A^{u\bar u}(Z') &=\frac{g}{2\sqrt{3}c_W} \sqrt{f(2/\sqrt{3})}
~\left[1+s_W^2\right]= 0.407\,,\\
\Delta_L^{\nu\bar\nu}(Z') & = \frac{g}{2\sqrt{3}c_W}\sqrt{f(2/\sqrt{3})} \left[1-3s_W^2\right]= 0.082 \,,\\
 \Delta_V^{\mu\bar\mu}(Z') &=\frac{g}{2\sqrt{3}c_W} \sqrt{f(2/\sqrt{3})}
~\left[1-7s_W^2\right] = -0.242
 \,,\\
 \Delta_A^{\mu\bar\mu}(Z') &=\frac{g}{2\sqrt{3}c_W} \sqrt{f(2/\sqrt{3})}
~\left[-1-s_W^2\right]= -0.407
\end{align}
\end{subequations}}%
 and for $\beta=-2/\sqrt{3}$
{\allowdisplaybreaks
\begin{subequations}
\begin{align}
\Delta_V^{d\bar d}(Z') &=\frac{g}{2\sqrt{3}c_W} \sqrt{f(2/\sqrt{3})}
~\left[-1+\frac{5}{3}s_W^2\right] = -0.191
 \,,\\
 \Delta_A^{d\bar d}(Z') &=\frac{g}{2\sqrt{3}c_W} \sqrt{f(2/\sqrt{3})}
~\left[1+s_W^2\right]= 0.407 \,,\\
 \Delta_V^{u\bar u}(Z') &=\frac{g}{2\sqrt{3}c_W} \sqrt{f(2/\sqrt{3})}
~\left[-1-\frac{7}{3}s_W^2\right] = -0.515
 \,,\\
 \Delta_A^{u\bar u}(Z') &=\frac{g}{2\sqrt{3}c_W} \sqrt{f(2/\sqrt{3})}
~\left[1-3s_W^2\right]= 0.082 \,,\\
\Delta_L^{\nu\bar\nu}(Z') & = \frac{g}{2\sqrt{3}c_W}\sqrt{f(2/\sqrt{3})} \left[1+s_W^2\right]= 0.407 \,,\\
 \Delta_V^{\mu\bar\mu}(Z') &=\frac{g}{2\sqrt{3}c_W} \sqrt{f(2/\sqrt{3})}
~\left[1+5s_W^2\right] = 0.731
 \,,\\
 \Delta_A^{\mu\bar\mu}(Z') &=\frac{g}{2\sqrt{3}c_W} \sqrt{f(2/\sqrt{3})}
~\left[-1+3s_W^2\right]= -0.082
\end{align}
\end{subequations}}%

These results confirm the ones seen in  Fig.~\ref{fig:F7}. For completeness we also list
the formulae for $\beta=\pm\sqrt{3}$ in order to demonstrate that for
$\beta=\sqrt{3}$ the couplings are too small to provide relevant NP effects,
while for $\beta=-\sqrt{3}$ they are too large to be consistent with the
flavour data and LEP-II bounds for $M_{Z^\prime}< 4\tev$, for which this model
is valid because of the Landau singularities in question. In order to stay
away from this singularity we give the values of couplings for $M_{Z^\prime}=2\tev$, that is for $\sin^2\theta_W=0.246$ and $g=0.636$.

\boldmath
\subsubsection*{$\beta=\pm\sqrt{3}$}
\unboldmath
For both signs we have
\be
 \Delta_L^{ij}(Z') =\frac{g}{\sqrt{3}}c_W \sqrt{f(\sqrt{3})} v_{3i}^*v_{3j} =
  ~ 2.52\, v_{3i}^*v_{3j}
\ee

In the case of $\beta=\sqrt{3}$ the formulae for leptonic couplings are modified \cite{Buras:2012dp}. We have then
{\allowdisplaybreaks
\begin{subequations}
\begin{align}
\Delta_V^{d\bar d}(Z') &=\frac{g}{2\sqrt{3}c_W} \sqrt{f\sqrt{3})}
~\left[-1\right] = -1.672
 \,,\\
 \Delta_A^{d\bar d}(Z') &=\frac{g}{2\sqrt{3}c_W} \sqrt{f(\sqrt{3})}
~\left[1-4s_W^2\right]= 0.027 \,,\\
 \Delta_V^{u\bar u}(Z') &=\frac{g}{2\sqrt{3}c_W} \sqrt{f(\sqrt{3})}
~\left[-1+6s_W^2\right] = 0.796
 \,,\\
 \Delta_A^{u\bar u}(Z') &=\frac{g}{2\sqrt{3}c_W} \sqrt{f(\sqrt{3})}
~\left[1+2s_W^2\right]= 2.494 \,,\\
\Delta_L^{\nu\bar\nu}(Z') & = \frac{g}{2\sqrt{3}c_W}\sqrt{f(\sqrt{3})} \left[1-4s_W^2\right]= 0.027 \,,\\
 \Delta_V^{\mu\bar\mu}(Z') &=\frac{3 g}{2\sqrt{3}c_W} \sqrt{f(\sqrt{3})}
~\left[1-4s_W^2\right] = 0.080
 \,,\\
 \Delta_A^{\mu\bar\mu}(Z') &=\frac{g}{2\sqrt{3}c_W} \sqrt{f(\sqrt{3})}
~\left[1-4 s_W^2\right]= 0.027
\end{align}
\end{subequations}}%
 and for $\beta=-\sqrt{3}$
{\allowdisplaybreaks
\begin{subequations}
\begin{align}
\Delta_V^{d\bar d}(Z') &=\frac{g}{2\sqrt{3}c_W} \sqrt{f\sqrt{3})}
~\left[-1+2s_W^2\right] = -0.849
 \,,\\
 \Delta_A^{d\bar d}(Z') &=\frac{g}{2\sqrt{3}c_W} \sqrt{f(\sqrt{3})}
~\left[1+2s_W^2\right]= 2.494 \,,\\
 \Delta_V^{u\bar u}(Z') &=\frac{g}{2\sqrt{3}c_W} \sqrt{f(\sqrt{3})}
~\left[-1-4s_W^2\right] = -3.316
 \,,\\
 \Delta_A^{u\bar u}(Z') &=\frac{g}{2\sqrt{3}c_W} \sqrt{f(\sqrt{3})}
~\left[1-4s_W^2\right]= 0.027 \,,\\
\Delta_L^{\nu\bar\nu}(Z') & = \frac{g}{2\sqrt{3}c_W}\sqrt{f(\sqrt{3})} \left[1+2s_W^2\right]= 2.49  \,,\\
 \Delta_V^{\mu\bar\mu}(Z') &=\frac{g}{2\sqrt{3}c_W} \sqrt{f(\sqrt{3})}
~\left[1+8s_W^2\right] = 4.96
 \,,\\
 \Delta_A^{\mu\bar\mu}(Z') &=\frac{g}{2\sqrt{3}c_W} \sqrt{f(\sqrt{3})}
~\left[-1+4s_W^2\right]= -0.027
\end{align}
\end{subequations}}%

\boldmath
\subsubsection*{SM couplings of $Z$}
\unboldmath
For comparison we give the couplings of $Z$ boson that we evaluate with
$g=0.652$ and $\sin^2\theta_W=0.23116$ as valid at $M_Z$. The non-diagonal
couplings vanish at tree-level and the diagonal ones are given as follows:

{\allowdisplaybreaks
\begin{subequations}
\begin{align}
\Delta_V^{d\bar d}(Z) &=\frac{g}{2 c_W}
~\left[-1+\frac{4}{3}s_W^2\right] = -0.257
 \,,\\
 \Delta_A^{d\bar d}(Z) &=\frac{g}{2 c_W} =  0.372 \,,\\
 \Delta_V^{u\bar u}(Z) &=\frac{g}{2 c_W}
~\left[1-\frac{8}{3}s_W^2\right] = 0.143
 \,,\\
 \Delta_A^{u\bar u}(Z) &=-\frac{g}{2 c_W} = -0.372 \,,\\
\Delta_L^{\nu\bar\nu}(Z) & = \frac{g}{2 c_W}= 0.372  \,,\\
 \Delta_V^{\mu\bar\mu}(Z) &=-\frac{g}{2 c_W}
~\left[1-4s_W^2\right] = -0.028
 \,,\\
 \Delta_A^{\mu\bar\mu}(Z) &=\frac{g}{2 c_W} = 0.372
\end{align}
\end{subequations}}%

\section{LEP-II constraints}\label{app:LEP-II}
 We will list here formulae which we used to verify that the four 331 models
investigated by us satisfy LEP-II constraints on $M_{Z^\prime}$. To this end
we generalized the usual SM expressions to include $Z^\prime$ contribution.
In this context we found the presentation in the book of Burgess and Moore
\cite{Burgess:2007zi}  useful.

The cross section for $e^+\e^-\to f\bar f$ where $f$ is a lepton or quark is
given in terms of helicity amplitudes $A^{ef}_{ij}$ by
\be
\sigma(e^+\e^-\to f\bar f)=\frac{\pi\alpha^2s N_c}{3}\left(|A^{ef}_{\rm LL}(s)|^2+
|A^{ef}_{\rm RR}(s)|^2+|A^{ef}_{\rm LR}(s)|^2+|A^{ef}_{\rm RL}(s)|^2\right),
\ee
where $N_c=3(1)$ for quarks (leptons).

For FB and LR asymmetries we have
\be
\mathcal{A}_{\rm FB}(e^+\e^-\to f\bar f)=\frac{3}{4}\frac{\left(|A^{ef}_{\rm LL}(s)|^2+
|A^{ef}_{\rm RR}(s)|^2-|A^{ef}_{\rm LR}(s)|^2-|A^{ef}_{\rm RL}(s)|^2\right)}{\left(|A^{ef}_{\rm LL}(s)|^2+
|A^{ef}_{\rm RR}(s)|^2+|A^{ef}_{\rm LR}(s)|^2+|A^{ef}_{\rm RL}(s)|^2\right)}
\ee

and

\be
\mathcal{A}_{\rm LR}(e^+\e^-\to f\bar f)=\frac{\left(|A^{ef}_{\rm LL}(s)|^2+
|A^{ef}_{\rm LR}(s)|^2-|A^{ef}_{\rm RR}(s)|^2-|A^{ef}_{\rm RL}(s)|^2\right)}{\left(|A^{ef}_{\rm LL}(s)|^2+
|A^{ef}_{\rm RR}(s)|^2+|A^{ef}_{\rm LR}(s)|^2+|A^{ef}_{\rm RL}(s)|^2\right)}.
\ee

The helicity amplitudes are given for
$M_{Z^\prime}\gg\sqrt{s}$  in the $Z^\prime$ models generally as follows (we drop
the argument $s$)
\be
A^{ef}_{ij}=A_{ij}^{\rm SM}+A_{ij}^{\rm NP}, \qquad i,j={\rm L,R}
\ee
Defining then
\be
\eta_{ij}^{ef}(Z^\prime)=-\Delta^{\e\bar\e}_i(Z^\prime)\Delta^{f\bar f}_j(Z^\prime),
\ee
where the minus sign comes from $Z^\prime$ propagator but
\be
\eta_{ij}^{ef}(Z)=\Delta^{\e\bar\e}_i(Z)\Delta^{f\bar f}_j(Z),
\ee
without this minus sign ($\sqrt{s}> M_Z$)
we have
\be
A_{ij}^{\rm SM}=\frac{Q_e Q_f}{s}+\frac{1}{4\pi\alpha}\left[\frac{\eta_{ij}^{ef}(Z)}{s-M_Z^2}\right], \qquad
A_{ij}^{\rm NP}=\frac{1}{4\pi\alpha}\left[\frac{\eta_{ij}^{ef}(Z^\prime)}{M_{Z^\prime}^2}\right].
\ee
Here the first term in the SM contribution represents photon contribution.
Note that for the values of $M_{Z^\prime}$ considered, $s$ in the $Z^\prime$
propagator can be neglected, while for $\sqrt{s}=200\gev$ one has
$\sqrt{s-M_Z^2}=178\gev$.

One can define the shift in the cross section due to NP contributions simply
as follows:
\be
\Delta\sigma^{\rm NP}=\sigma(e^+\e^-\to f\bar f)-\sigma^{\rm SM}.
\ee
In view of small NP effects only the interference between NP and SM matters and
we find
\be
\sigma^{\rm SM}=\frac{\pi\alpha^2s N_c}{3}\left(|A^{\rm SM}_{\rm LL}|^2+
|A^{\rm SM}_{\rm RR}|^2+|A^{\rm SM}_{\rm LR}|^2+|A^{\rm SM}_{\rm RL}|^2\right)\ee
and
\be
\Delta\sigma^{\rm NP}=
2\frac{\pi\alpha^2sN_c}{3}\left(A^{\rm SM}_{\rm LL}A^{\rm NP}_{\rm LL}+
A^{\rm SM}_{\rm RR}A^{\rm NP}_{\rm RR}+A^{\rm SM}_{\rm LR}A^{\rm NP}_{\rm LR}+
A^{\rm SM}_{\rm RL}A^{\rm NP}_{\rm RL}\right).
\ee

Analogous formulae can be derived for corrections to FB and LR asymmetries.

\boldmath
\section{The $\beta=\pm\sqrt{3}$ models}\label{beta3}
\unboldmath
Here we list the problems of $\beta=\pm\sqrt{3}$ models which originate 
in the value of the coupling $g_X$ which is not free but for a fixed $\beta$ 
is given in terms of $g$ and $\sin^2\theta_W$ as follows:
\be\label{gX}
{g_X^2}=g^2\frac{6 \sin^2 \theta_W}{1-(1+\beta^2) \sin^2 \theta_W}.
\ee

This formula implies for $\beta=\pm\sqrt{3}$
 a Landau singularity for $\sin^2\theta_W=0.25$ and this value is
reached through the renormalization group evolution of the SM couplings
for $M_{Z^\prime}$ typically around  $4\tev$ \cite{Ng:1992st,Frampton:2002st}\footnote{In fact we confirmed Frampton's result that at
one-loop level the singularity is reached precisely at  $M_{Z^\prime}=4\tev$ and this result is practically
unchanged at NLO. We thank David Straub for checking this.}. Therefore these models as they stand, even if  $V_L\not=V_{\rm CKM}$, can only
be valid for $M_{Z^\prime} < 4\tev$. Although in principle some new dynamics entering
around these scales could shift the Landau singularity to higher scales,
in particular supersymmetry \cite{Dias:2004wk,Dias:2004dc}, one should 
realize that even at $\mu=80\gev$ the coupling would be as large as 
 $\alpha_X\approx 0.6$ that is 
much larger than all couplings of the SM. 
At the relevant scales of order few TeV $\alpha_X\ge 2.5$  implying 
that  perturbative calculations cannot be trusted even in the presence of a
large $M_{Z^\prime}$. This is not the 
problem for  other four models discussed by us, where at $\mu=3\tev$, 
the coupling $\alpha_X$ equals approximately $0.07$ and $0.11$ for 
$\beta=\pm 1/\sqrt{3}$ and $\beta=\pm 2/\sqrt{3}$, respectively.

The related problems are as follows
\begin{itemize}
\item
Noting that the masses of the new charged gauge bosons $V$ and $Y$ are related within
$1\%$ accuracy to $M_{Z^\prime}$ through
\be
M_V=M_Y=M_{Z^\prime}\sqrt{1-(1+\beta^2)s_W^2},
\ee
we find  for $|\beta|=\sqrt{3}$ and $s_W^2=0.24-0.25$, valid for  $M_{Z^\prime}$ in the
ballpark of a few TeV, the masses of other heavy gauge bosons  $M_V=M_Y\le M_{Z^\prime}/5$. This is basically ruled out
by the LHC for  $M_{Z^\prime}\le 4\tev$. However, a dedicated study would be necessary in order to put this statement on the firm footing. This is not a problem for $|\beta|=1/\sqrt{3}$ and  $|\beta|=2/\sqrt{3}$, where we find
$M_V=M_Y\approx 0.8\, M_{Z^\prime}$
and
$M_V=M_Y\approx 0.7\, M_{Z^\prime}$, respectively.
\item
With the matrix $V_L$ equal to the CKM matrix we find that even for
values of  $M_{Z^\prime}=(5-7)\tev$ as considered in \cite{Gauld:2013qja} the
mass differences $\Delta M_s$ and $\Delta M_d$ are enhanced at least by a factor of two ($C_{B_{s,d}}\approx 2$)
relative to the SM values. In our view  it is unlikely that the future lattice
values of $\sqrt{\hat B_{B_s}}F_{B_s}$ and $\sqrt{\hat B_{B_d}}F_{B_d}$ would
change so much  to allow for a satisfactory description of the data for $\Delta M_{s,d}$  in this model. While choosing
$V_L\not =V_{\rm CKM}$ would remove this problem, this
does not help because of the last difficulty.
\item
It turns out that the size of predicted coupling
$\Delta_V^{\mu\bar\mu}(Z^\prime)$ in
this model implies through LEP-II data
a lower bound on $M_{Z^\prime}$  of order of $10\tev$ when
RG effects in $\sin^2\theta_W$ are taken into account\footnote{We thank Francois Richard for pointing out the  inconsistency of
the model in \cite{Gauld:2013qja}
 with the LEP-II data even in the absence of RG effects. We refer to his analysis of 331 models in \cite{Richard:2013xfa}, where the prospects for testing 331 models at the future ILC are presented.}. This  value is
 outside the validity of the model unless complicated new dynamics is introduced at scales of few TeV. 
\end{itemize}

\bibliographystyle{JHEP}
\bibliography{allrefs}
\end{document}